\documentclass[fleqn,usenatbib]{mnras}

\usepackage[T1]{fontenc}

\DeclareRobustCommand{\VAN}[3]{#2}
\let\VANthebibliography\thebibliography
\def\thebibliography{\DeclareRobustCommand{\VAN}[3]{##3}\VANthebibliography}

\usepackage{graphicx}	
\usepackage{amsmath}	
\usepackage{amssymb}	
\usepackage{siunitx}    
\usepackage{threeparttable}  

\newcommand{\mps}[1]{\SI{#1}{\meter\per\second}}
\renewcommand{\ion}[2]{{#1}\,{\small\sc #2}}
\newcommand{\atom}[2]{$^{#2}${#1}}
\newcommand{\daa}{\Delta\alpha/\alpha}
\newcommand{\figref}[1]{Fig.~\ref{#1}}
\newcommand{\secref}[1]{Section~\ref{#1}}
\newcommand{\tabref}[1]{Table~\ref{#1}}
\newcommand{\vpfit}{\texttt{VPFIT}}
\newcommand{\mvpfit}{\texttt{AI-VPFIT}}

\title[A new era of $\alpha$ measurements at high $z$]{A new era of fine structure constant measurements at high redshift}

\author[D. Milakovi{\'c} et al.]{Dinko Milakovi{\'c},$^{1}$\thanks{E-mail: dmilakov@protonmail.com}
Chung-Chi Lee,$^{2}$
Robert F. Carswell,$^{3}$
John K. Webb,$^{4}$ \newauthor
 Paolo Molaro,$^{5}$
and Luca Pasquini$^{1}$
\\
$^{1}${European Southern Observatory, Karl-Schwarzschild-str 2, 85748 Garching bei M{\"u}nchen, Germany}\\
$^{2}${DAMTP, Centre for Mathematical Sciences, University of Cambridge, Cambridge CB3 0WA, UK}\\
$^{3}${Institute of Astronomy, Madingley Road, Cambridge CB3 0HA, U.K.}\\
$^{4}${School of Physics, University of New South Wales, Sydney NSW 2052, Australia}\\
$^{5}${INAF-Osservatorio Astronomico di Trieste, Via G.B. Tiepolo 11, I-34143 Trieste, Italy}
}
\date{Accepted XXX. Received YYY; in original form ZZZ}

\pubyear{2020}
\usepackage{newtxtext,newtxmath}

\begin{document}
\label{firstpage}
\pagerange{\pageref{firstpage}--\pageref{lastpage}}
\maketitle

\begin{abstract}
New observations of the quasar HE0515$-$4414 have been made, aided by the Laser Frequency Comb (LFC), using the HARPS spectrograph on the ESO 3.6m telescope. We present three important advances for $\alpha$ measurements in quasar absorption spectra from these observations. Firstly, the data have been wavelength calibrated using LFC and ThAr methods. The LFC wavelength calibration residuals are six times smaller than when using the standard ThAr calibration. We give a direct comparison between $\alpha$ measurements made using the two methods. Secondly, spectral modelling was performed using Artificial Intelligence (fully automated, all human bias eliminated), including a temperature parameter for each absorption component. Thirdly, in contrast to previous work, additional model parameters were assigned to measure $\alpha$ for each individual absorption component. The increase in statistical uncertainty from the larger number of model parameters is small and the method allows a substantial advantage; outliers that would otherwise contribute a significant systematic, possibly corrupting the entire measurement, are identified and removed, permitting a more robust overall result. The $z_{abs} \!\! = \!\! 1.15$ absorption system along the HE0515$-$4414 sightline yields 40 new $\alpha$ measurements. We constrain spatial fluctuations in $\alpha$ to be $\daa \leq 9 \times 10^{-5}$ on scales $\approx \!\! \SI{20}{\kilo\meter\per\second}$, corresponding to $\approx25\;$kpc if the $z_{abs} \!\! = \!\! 1.15$ system arises in a 1Mpc cluster. Collectively, the 40 measurements yield $\daa=-0.27\pm2.41\times10^{-6}$, consistent with no variation.
\end{abstract}




\section{Introduction}
Fundamental constants, such as the fine structure constant ($\alpha\equiv\frac{1}{4\pi\epsilon_0}\frac{e^2}{\hbar c}$) and the proton-to-electron mass ratio ($\mu \equiv \frac{m_p}{m_e}$), are expected to vary in some modifications of General Relativity. A scalar field $\phi$ coupling to the baryonic matter can produce temporal and/or spatial $\alpha$ variations  \citep{Bekenstein1982,Sandvik2002,Shaw2005,Barrow2012,Copeland2004,Marra2005}. $\alpha$ may also vary with gravitational potential \citep{Dicke1959,Sandvik2002,Mota2004PLB,Mota2004MNRAS}, or via interactions of baryonic matter with dark matter candidates \citep{Olive2002,Stadnik2015}, or if the vacuum expectation value of $\phi$ depends on the local density \citep{Silva2014}. In theories with extra spatial dimensions (e.g. Kaluza-Klein and string theories), expansion (or contraction) of higher dimensions can produce observed changes to the coupling constants in our 4-dimensional space time. Recent reviews of varying constants are given by \citet{Uzan2011,Martins2017}.

Variations in $\alpha$ and $\mu$ have been explored both on Earth through atomic clock measurements \citep{Rosenband2008}, isotope ratio studies \citep{Damour1996}, and in space using astronomical observations of white dwarfs \citep{Berengut2013,Bainbridge2017Univ}, galaxies \citep{Bahcall2004}, quasars \citep{Webb1999,Murphy2003,Wilczynska2015,Ubachs2018}, stars around the supermassive black hole in the Galaxy \citep{Hees2020}, and the Cosmic Microwave Background \citep{Avelino2001,Planck2015}. A comprehensive analysis of 317 quasar absorption systems using the {\it Many Multiplet method} \citep{Dzuba1999,Webb1999} hinted at a spatial variation of $\alpha$, modelled as a dipole with amplitude $\Delta\alpha/\alpha = (\alpha_Q-\alpha_0)/\alpha_0=1.1\pm0.2\times10^{-6}$, where $\alpha_Q$ are quasar absorption measurements and $\alpha_0$ is the terrestrial value \citep{Webb2011,King2012,Wilczynska2020}. 

Echelle spectrographs, using slit-based observations and calibrated using ThAr methods, are prone to long-range wavelength distortions \citep{Molaro2008,Rahmani2013,Evans2014}. Such distortions, if present and left uncorrected, can significantly contribute to the total $\daa$ measurement uncertainty \citep{Evans2014,Kotus2017}. Correction techniques include using additional external calibration information from asteroid observations \citep{Molaro2013,Rahmani2013}, iodine cells \citep{Griest2010,Whitmore2010}, solar-twin observations \citep{Whitmore2015}, or by using additional model parameters \citep{Dumont2017}. To a reasonable approximation, the best wavelength correction that could be achieved with any of these methods has an accuracy no better than $\lesssim \SI{30}{\meter\per\second}$. For comparison, the best laboratory accuracy of UV wavelengths used for $\daa$ measurements is \SI{0.01}{\meter\per\second}, three orders of magnitude better. Laser Frequency Comb \citep[LFC,][]{Udem2002,Haensch2006,Steinmetz2008} wavelength calibration methods provide a vastly superior calibration than the correction methods above as they provide \SI{3}{\meter\per\second} accuracy for individual line center measurements \citep{Probst2020,Milakovic2020}. 

In this paper, we report a set of high redshift $\alpha$ measurements from new observations of the quasar HE0515$-$4414. The observations (described in \secref{sec:data}) are of very high quality. These data are the first quasar spectral observations where the wavelength calibration has been carried out using an LFC.  This means that any wavelength scale distortions present will be negligible. A second spectrum was produced from the same quasar observations but calibrated using ThAr methods. The two spectra enable a unique set of comparative tests to quantify uncertainties in searches for fundamental constant variations.

We use new automated analysis methods \citep{Lee2020} to produce models for each spectrum and measure $\alpha$ using the Many Multiplet method (\secref{sec:modelling}). We introduce a new method, measuring $\alpha$ for each absorption component (rather than an average across an entire absorption complex). This provides considerably more detail and also offers a substantial advantage by enabling systematics to be more readily identified. We summarise our main findings in \secref{sec:results} and discuss them in \secref{sec:discussion}.

\section{Data}
\label{sec:data}
\subsection{Data acquisition}
\label{sec:data_acquisition}
The spectrum used in this work was produced from high-resolution ($R=\frac{\lambda}{\Delta\lambda}=115000$) observations using the High Accuracy Radial velocity Planet Searcher (HARPS) echelle spectrograph \citep{Mayor2003}. HARPS is a double-channel echelle spectrograph built for extremely precise spectroscopic measurements. 
We observed HE0515$-$0414 (abbreviated as HE0515) between $3^{\rm rd}$ and $11^{\rm th}$ December 2018 using HARPS in the classical fibre spectroscopy mode, where channel A recorded the HE0515 spectrum and channel B recorded the sky spectrum. We obtained 36 exposures totalling 52h 31m (Table~\ref{tab:observations}). Each exposure was bracketed by ThAr and LFC exposures for wavelength calibration. The sky was dark and the seeing conditions varied between 0.45 and 1.98 arcsec throughout the observing run. The median seeing (i.e.\ the median of column 3 in Table~\ref{tab:observations}) is 1.34 arcsec. This has no influence on the final spectral resolution. The secondary guiding system ensures the object is consistently centered on the object image up to 0.01 arcsec and octagonal fibres ensure that the light evenly illuminates the spectrograph pupil. Therefore, telescope guiding and fibre illumination are not expected to introduce spectroscopic velocity shifts larger than \mps{0.12}~ \citep{LoCurto2015}.

Light entering the spectrograph is recorded on the detector (a mosaic of two EEV2k4 CCDs) for which the read-out mechanism is located on one of its sides \citep{Mayor2003,HARPS-manual}. By design, charge transfer occurs in the cross-dispersion direction to minimise effects of charge transfer inefficiency (CTI). Left uncorrected, CTI can introduce spurious spectroscopic velocity shifts up to \mps{3} for very low flux exposures \citep{Milakovic2020}. However, as no appropriate CTI model yet exists for HARPS, we do not correct for this effect. 

\begin{table}
    \centering
    \caption{The final co-added spectrum of HE0515$-$4414 is formed from co-adding 36 HARPS exposures taken in classic spectroscopy mode, totalling 52h 31m. Columns 1 and 2 give the observing time start (in UTC) and the exposure time, respectively. Column 3 gives the average of the telescope seeing recorded at the beginning and the end of observation. Column 4 gives the S/N per extracted pixel at the center of order 111 ($\approx\SI{5500}{\angstrom}$). All quantities are determined from values recorded in headers of \texttt{e2ds} HARPS pipeline products.}
    \label{tab:observations}
    \vspace{0.3cm}
    \begin{tabular}{cccc}
        \hline
          Observing time & Exp. time & Seeing & S/N   \\
          (UTC) & (s) & (arcsec) & (${\rm pix^{-1}}$) \\
        \hline\hline
         2018-12-04T00:27:52.031 & 5400 & 1.48 & 6.4 \\
         2018-12-04T02:12:04.582 & 5400 & 1.35 & 11.0 \\
         2018-12-04T03:50:40.736 & 5400 & 1.54 & 8.3 \\
         2018-12-04T05:36:27.032 & 5400 & 1.71 & 5.6 \\
         2018-12-04T07:14:20.953 & 2700 & 1.25 & 3.2 \\
         2018-12-05T03:14:39.850 & 5400 & 1.91 & 7.3 \\
         2018-12-05T04:52:04.530 & 5400 & 0.45 & 7.1 \\
         2018-12-05T06:42:00.250 & 5400 & N/A  & 7.5 \\
         2018-12-06T00:41:08.634 & 5400 & 1.98 & 5.8 \\
         2018-12-06T02:30:00.882 & 5400 & 1.64 & 9.0 \\
         2018-12-06T04:08:04.226 & 5400 & 1.40 & 4.6 \\
         2018-12-06T05:46:25.189 & 5400 & 1.58 & 8.2 \\
         2018-12-06T07:25:04.433 & 5098 & 1.59 & 6.4 \\
         2018-12-07T00:23:24.209 & 5400 & 1.50 & 7.1 \\
         2018-12-07T02:01:11.070 & 5400 & 1.33 & 9.1 \\
         2018-12-07T03:38:28.641 & 4905 & 1.66 & 7.2 \\
         2018-12-07T05:32:14.425 & 5400 & 1.31 & 7.9 \\
         2018-12-07T07:09:39.678 & 5400 & 1.40 & 5.3 \\
         2018-12-08T00:32:12.597 & 4214 & 1.32 & 5.0 \\
         2018-12-08T02:15:25.854 & 5400 & 1.44 & 5.9 \\
         2018-12-08T03:54:16.167 & 5400 & 1.32 & 9.0 \\
         2018-12-08T05:32:25.299 & 5400 & 1.39 & 11.4 \\
         2018-12-08T07:10:06.569 & 5400 & 1.26 & 9.6 \\
         2018-12-09T00:37:46.416 & 5400 & 1.29 & 8.2 \\
         2018-12-09T02:22:05.279 & 5400 & 1.21 & 10.1 \\
         2018-12-09T03:59:32.258 & 5400 & 1.17 & 9.3 \\
         2018-12-09T05:36:04.415 & 5400 & 1.10 & 11.0 \\
         2018-12-10T00:24:18.778 & 5400 & 1.30 & 9.2 \\
         2018-12-10T02:15:06.228 & 5400 & 1.69 & 9.1 \\
         2018-12-10T03:54:46.885 & 5400 & 1.23 & 8.6 \\
         2018-12-10T05:32:28.955 & 5400 & 1.07 & 10.6 \\
         2018-12-10T07:10:20.256 & 5400 & 0.80 & 13.2 \\
         2018-12-11T00:28:14.801 & 5400 & 1.04 & 8.9 \\
         2018-12-11T02:12:13.372 & 5400 & 1.56 & 11.2 \\
         2018-12-11T03:50:37.414 & 5400 & 1.27 & 10.7 \\
         2018-12-11T07:31:21.735 & 4795 & 1.05 & 8.6 \\
        \hline
    \end{tabular}

\end{table}
Raw images were reduced using the HARPS pipeline \citep[version 3.8,][]{HARPS-manual}. The pipeline extracts 1d spectra of individual echelle orders following optimal extraction by \cite{Horne1986,Robertson1986}. Order tracing and pixel weights are determined from tungsten-lamp frames taken at the beginning of each night. Pipeline products previously demonstrated a \mps{0.01} precision \citep{Milakovic2020}, so we do not expect spectroscopic velocity shifts associated with its use. 

\subsection{Wavelength calibration and data addition}
\label{sec:data_calibration}
Wavelength calibration was obtained from LFC and ThAr frames taken immediately before each quasar exposure. The LFC has an offset frequency of \SI{4.58}{\giga\hertz} and \SI{18}{\giga\hertz} line separation. LFC wavelength calibration was performed using eight 7$^{\rm th}$ order polynomials per echelle order. Each echelle order spans eight 512-pixel blocks on the CCD \citep{Wilken2010,Molaro2013}. The accuracy of the LFC wavelength calibration is \mps{3}, measured by the root-mean-square (rms) of calibration residuals \citep[i.e.\ known LFC line frequency minus the frequency determined from the wavelength solution at line position on the detector, see][]{Milakovic2020}. The ThAr wavelength calibration was produced by the HARPS pipeline using a single third order polynomial per echelle order, with an accuracy of \mps{27}. The average difference in the two calibrations, considering wavelengths $\lambda\geqslant \SI{5000}{\angstrom}$, is \mps{-1.13} (LFC minus ThAr). 

\begin{figure}
    \centering
    \includegraphics[width=\linewidth]{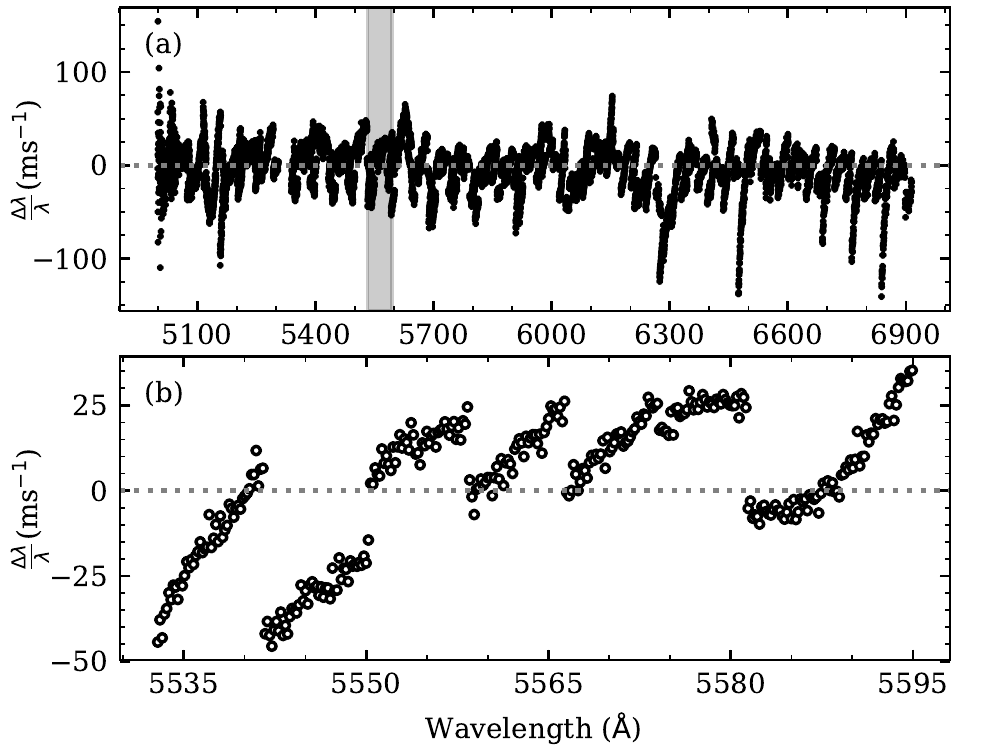}
    \caption{Distortions in the ThAr calibration revealed by comparing with LFC lines. Each dot is a single LFC line. Distortion amplitude generally increases at order edges. Clear discontinuities associated with the HARPS detector stitching pattern \citep{Wilken2010,Molaro2013} are seen for all orders. The distortions do not show long-range wavelength dependency, have a \mps{-1.1} mean offset with respect to the LFC scale, and an overall scatter of \mps{27.9} rms. Panel (b) shows the region covered by echelle order 110 ($\lambda\approx\SI{5500}{\angstrom}$), also marked by a grey rectangle in panel (a).}
    \label{fig:calibration_comparison}
\end{figure}

Comparing the true LFC line wavelengths to the ThAr-calibrated wavelengths at their location on the detector reveals a distortion pattern in the ThAr calibration, illustrated on \figref{fig:calibration_comparison}. The pattern shows no long-range wavelength trends, but contains discontinuities associated with stitching of the HARPS detector \citep{Wilken2010,Molaro2013}, not accounted for by the pipeline calibration procedure. We discuss the impact of this distortion pattern on $\alpha$ measurements in \secref{sec:discussion}. 

\begin{figure}
    \centering
    \includegraphics[width=\linewidth]{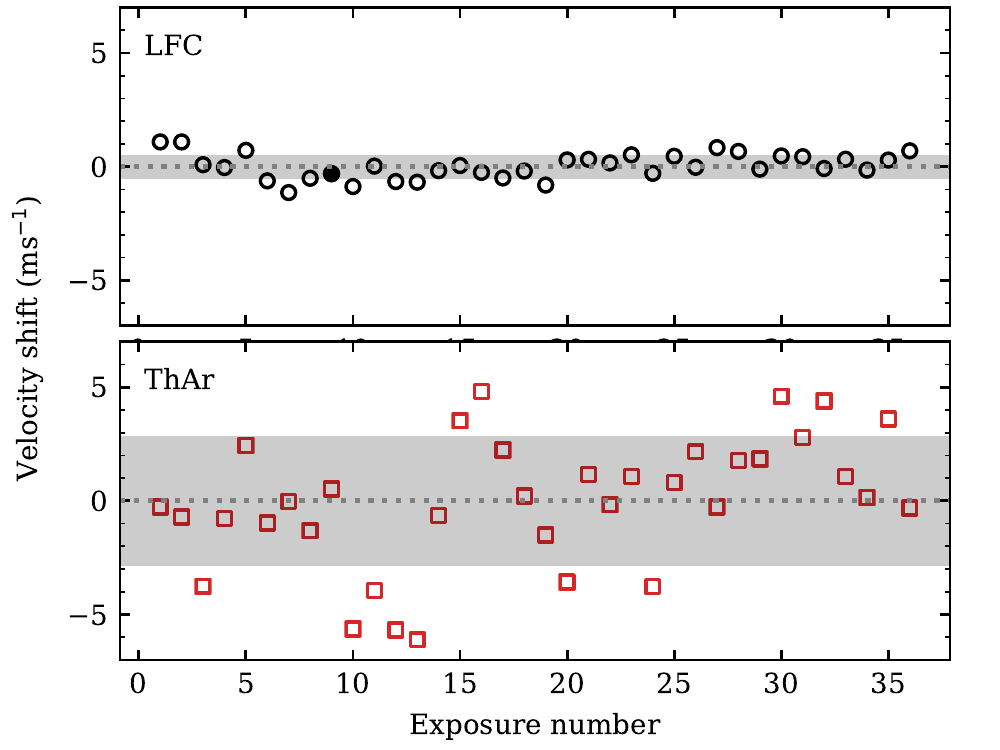}
    \caption{LFC calibration was found to drift by \mps{0.52} rms over the 8 nights of our run (grey shaded region in the top panel), as measured by shifts of LFC lines on the detector. Over the same period, ThAr calibration experienced drifts of \mps{2.87} rms (grey shaded region in the bottom panel). Shifts in individual LFC (ThAr) exposures are shown as unfilled black circles (red squares) in the top (bottom) panel. The zero line represents the mean value of all points. Shifts were calculated using echelle orders 88 to 121 only (those orders best covered by LFC). The filled black circle in the top panel represents the LFC exposure used for wavelength calibration of all quasar exposures (see text).}
    \label{fig:calibration_stability}
\end{figure}

Over the 8 nights of our run, the spectrograph stability is \mps{0.52}, as illustrated in \figref{fig:calibration_stability}\footnote{This is not the same as the precision which can be achieved in the simultaneous referencing observing mode.}. This number was obtained by measuring the average shifts of LFC line positions in individual exposures with respect to their position in the first exposure and calculating the rms. Applying the same method to the ThAr lines gives an rms of \mps{2.87}, six times larger.

Although all LFC exposures were taken under the same nominal conditions and with same exposure times, it turned out that one exposure was substantially better (in terms of flux) than all others. Therefore, after careful consistency checking between multiple LFC exposures, this highest flux LFC exposure was used to wavelength calibrate all quasar exposures for wavelengths $\lambda\geqslant\SI{5000}{\angstrom}$ (the LFC data cuts off below this wavelength). There are no saturated LFC lines. We do not follow the same procedure for ThAr calibration, but calibrate each quasar exposure using the ThAr frame taken immediately beforehand.

\begin{figure}
    \centering
    \includegraphics[width=\linewidth]{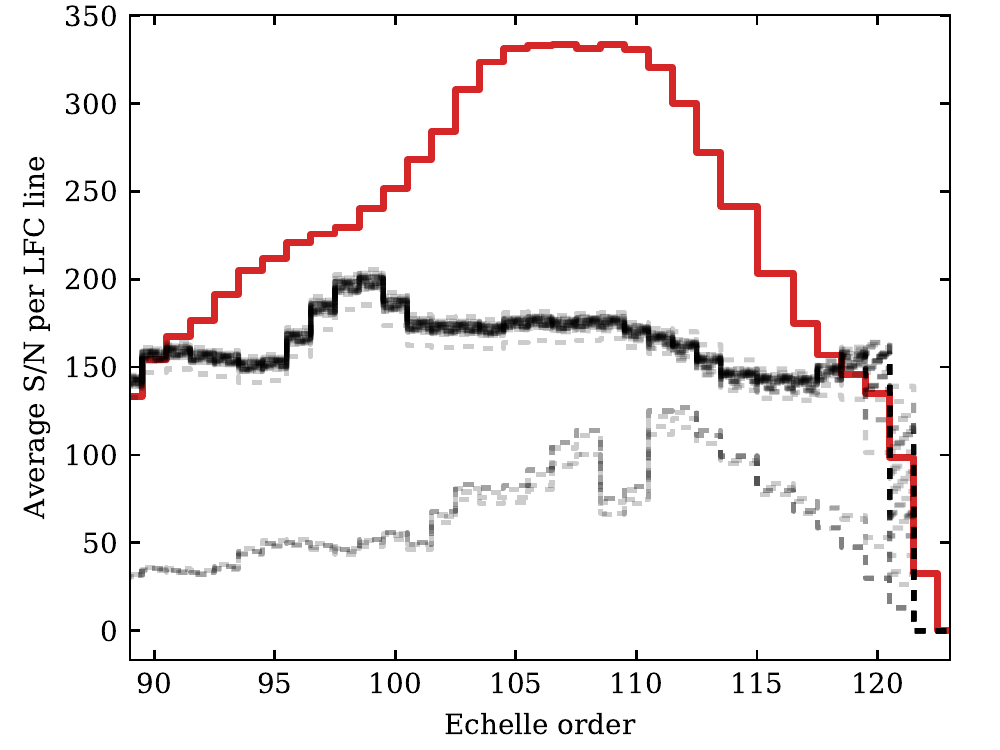}
    \caption{Each histogram shows the average S/N per LFC line in a single LFC exposure as a function of echelle order. Five exposures have significantly lower S/N than others (the reason is unknown). 
    A single exposure (full red histogram) reaches S/N of $\approx 330$ per LFC line and covers one additional echelle order (122). 
    This exposure provides the most accurate wavelength calibration over the broadest wavelength range. Given the exquisite stability of HARPS (see \figref{fig:calibration_stability}), we choose to use this exposure to calibrate all quasar exposures.}
    \label{fig:lfc_snr_per_order}
\end{figure}

We transform the LFC and the ThAr wavelength scales to the Solar system barycentre rest-frame using the barycentric velocity shift correction provided by the HARPS pipeline, independently for each quasar exposure. The barycentric correction provided by the pipeline is based on \citet{Bretagnon1988} and uses the flux-weighted average time of observation. This value agrees down to several \SI{}{\milli\meter\per\second} with our independent calculation, using the same information and the \texttt{astropy} module\footnote{https://docs.astropy.org/en/stable/coordinates/velocities.html}. 

\begin{figure*}
    \centering
    \includegraphics[width=\textwidth]{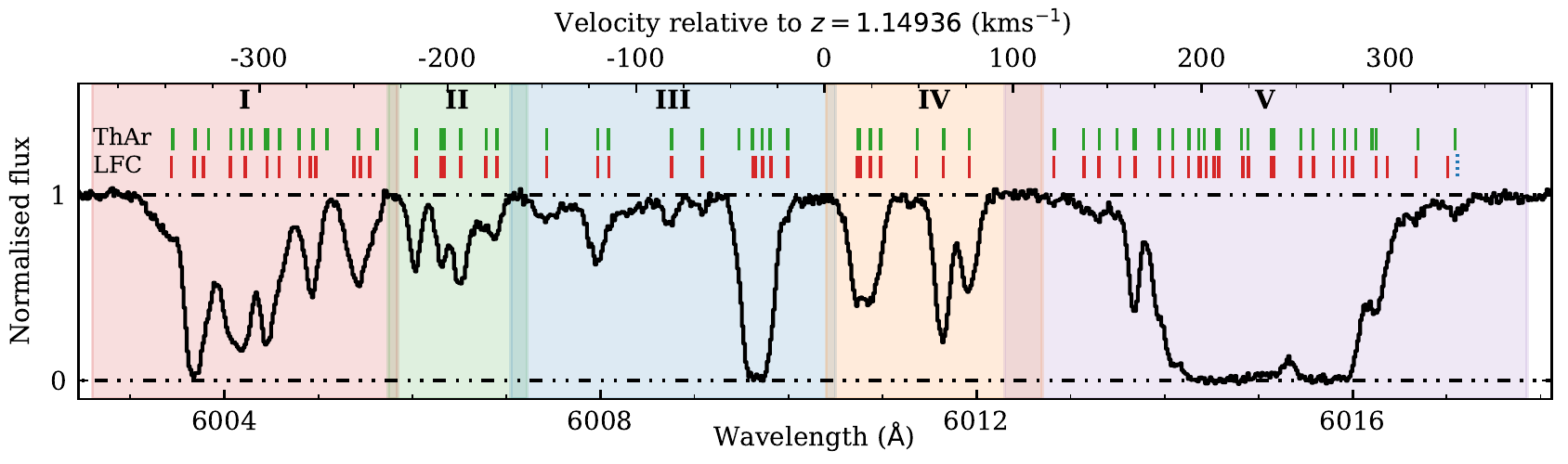}
    \caption{The LFC-calibrated spectrum of HE0515$-$4414 showing the \ion{Mg}{ii} $\lambda 2796$ transition at $z\approx1.15$ (black histogram). Five coloured areas mark individual regions (denoted by I-V) for which we produce \mvpfit~models and measure $\daa$. Small overlaps between neighbouring regions enables better continuum level estimation in each region. The solid red (green) ticks above the data indicate individual metal absorption lines in the best-fit LFC (ThAr) model. The top $x$-axis shows the velocity with respect to the average redshift of all metal lines in the LFC model, $z=1.14936$. The blue dashed line at $v\approx\SI{330}{\kilo\meter\per\second}$ marks the location of an unidentified absorption line in the LFC model. More detailed plots, showing the data, the model, and the residuals can be found in Appendix~\ref{sec:mvpfit_lfc_models} (LFC) and Appendix~\ref{sec:mvpfit_thar_models} (ThAr), split by region and transition.
}
    \label{fig:spec}
\end{figure*}
Finally, we rebin the individual extracted spectra onto a common wavelength grid using a custom routine and sum them together, weighting each pixel by its error estimate (which includes the Poissonian error term, the read-out noise, and the dark current). The error array extracted during this procedure agrees with the estimate derived from flux rms over $\approx\SI{1}{\angstrom}$ range. The final co-added spectrum has an average signal-to-noise ratio (S/N) near 50 per \SI{0.015}{\angstrom} pixel in the continuum. This data extraction process was performed for the LFC and ThAr calibration separately, producing two spectra from the same observations.

\section{Modelling procedure} 
\label{sec:modelling}
The spectrum shows a damped Lyman-$\alpha$ absorption complex spanning \SI{700}{\kilo\meter\per\second}, at redshift $z_{abs}\approx 1.15$ \citep{Reimers1998}, from which numerous previous measurements of $\alpha$ have been made \citep{Quast2004,Levshakov2005,Levshakov2006,Chand2006,Molaro2008,Kotus2017}. There are at least twenty-six transitions useful for an $\alpha$ measurement in this system. The {\it Many Multiplet} analysis in this work makes use of transitions covered by the LFC calibration, listed in \tabref{tab:transitions}. None of the transitions we use blend with any other systems nor with transitions from the $z=0.28$ absorption complex identified by \citet{Bielby2017}. The LFC-calibrated spectrum showing the \ion{Mg}{ii} $\lambda 2796$ transition is plotted as a black histogram in \figref{fig:spec}. 

We use the most recent set of laboratory wavelength measurements, transition probabilities, oscillator strengths, and isotopic structures for the relevant transitions. These are given in \tabref{tab:transitions}. The isotopic abundances were assumed to be solar \citep{Asplund2009}. The sensitivity coefficients that relate atomic line shifts to a change in $\alpha$ are from \cite{Dzuba1999b,Dzuba2002,Dzuba2009}. All atomic data is provided as online supplementary material.

We use a fully automated modelling procedure, \mvpfit , to produce a model of the absorption system \citep{Lee2020}. \mvpfit~is a development of the approach introduced in \citet{Bainbridge2017} and \citet{Bainbridge2017Univ}. Model complexity is increased by placing absorption components (``trial lines'') at a random location in the velocity structure and checking if the newly introduced parameters are justified by the data. For the analysis described in the present paper, the optimal number of model parameters are derived using the corrected Akaike Infomation Criterion \citep[AICc][]{Akaike1974,Sugiura1978}. Performance tests using simulated data are described in \citet{Lee2020}. Redshifts and $b$-parameters of components appearing in multiple species are tied during fitting (\secref{sec:temperature}). Column densities are free parameters. We include additional parameters for the unabsorbed continuum level for all transitions and for zero-level adjustment for saturated ones. $\alpha$ is also kept as a free parameter but this has been treated in two different ways (see Sections \ref{sec:5alpha} and \ref{sec:manyalpha}). The basis of the AI algorithm is a genetic process in which a model is built up in 6 well-defined stages. An initial model for the absorption system is generated using a ``primary'' species, that is, one atomic transition (or atomic species), selected to maximise line strength but avoiding line saturation. Subsequent stages incorporate further atomic species, with appropriately tied parameters, refine parameter errors, check for overfitting, and allow for ``interlopers'', i.e. unidentified lines from other redshift systems that are needed to derive a statistically acceptable overall model. 

We produce models for the LFC-calibrated and ThAr-calibrated spectrum independently. All relevant settings during \mvpfit~modelling are kept the same (such as the number of attempts \mvpfit~will make to increase model complexity before proceeding to the following stage, default parameter values for trial lines, line dropping criteria, finite derivative step sizes, etc.), ensuring that all the differences between the final models are a direct consequence of differences in the input data. We refer to models produced from the LFC- and the ThAr-calibrated spectrum as the LFC and the ThAr models, respectively. Figures showing the data, the models, and the residuals for all transitions and all regions are in Appendices~\ref{sec:mvpfit_lfc_models} (for LFC) and \ref{sec:mvpfit_thar_models} (for ThAr). 
\begin{figure*}
    \centering
    \includegraphics[width=\textwidth]{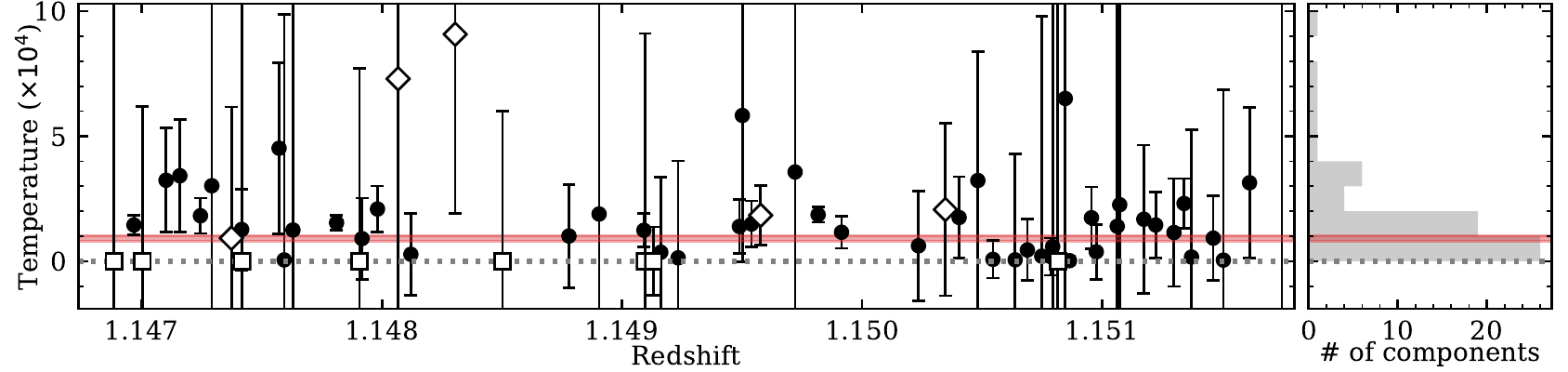}
    \caption{Temperatures of individual absorption components. Error bars represent $1\sigma$ uncertainties from the covariance matrix at the best-fit solution. Black filled circles show components having both thermal and turbulent broadening. Thermal-only fits are marked by white diamonds. Turbulent-only fits are shown as white squares. The red shaded area marks the weighted average temperature, $T=9.12\pm1.08\times\SI{e3}{\kelvin}$. Individual measurements are scattered around this value with $\chi^2_\nu=1.042$, where $\nu=61$. The panel to the right is the histogram of temperatures.
    }
    \label{fig:temp_vs_z}
\end{figure*}

\subsection{Instrumental profile}
The nominal HARPS instrumental profile has a FWHM of 2.61 km/s. The average $b$-parameter for individual absorption components in the absorption system analysed in this paper is $\approx 5$ km/s. The observed quasar lines are thus well-resolved. When matching models to the observed data we must convolve Voigt profiles with the HARPS instrumental profile (IP). To do this we used a Gaussian IP. However, slight departures from a Gaussian have been reported. Moreover, these are found to vary with both flux and position on the detector \citet{Milakovic2020}. A numerical profile determined directly from HARPS calibration data would thus provide a slightly more accurate IP. However, this was not possible due to insufficient available data\footnote{The IP is known to be flux-dependent. We do not have a suitable set of LFC exposures to determine the IP as a function of flux level. Data from \citet{Probst2020} and \citet{Milakovic2020} is not useful for this purpose because the HARPS fibres were exchanged since that data were taken, thus changing the IP \citep{LoCurto2015}}.

\subsection{Temperature as a free parameter}
\label{sec:temperature}
The choice of a line broadening mechanism heavily influences the final model. We found that using turbulent broadening (i.e.\ not including temperature as a free parameter) impacts significantly on the analysis. For example, imposing a turbulent model forces $b$ to be the same for all species irrespective of atomic mass. If turbulent broadening does not apply in practice, the consequence of the assumption is that additional velocity components are unavoidably included in order to achieve a satisfactory fit to the data. The converse is true - i.e.\ if a pure thermal model is imposed in the modelling procedure, additional velocity components may also be required to compensate if the model is inappropriate. We explored this by computing models for all three cases, i.e.\ turbulent, thermal, and mixed-$b$. 

In the mixed-$b$ model, the total line $b$-parameter is:
\begin{equation}
    \label{eq:btot}
    b^2 = b_{\rm turb}^2 + \frac{2kT}{m},
\end{equation}
where the right-hand-side terms are the turbulent and thermal contributions, respectively. In the thermal contribution, $k$ is the Boltzmann's constant, $T$ is gas temperature, and $m$ is the appropriate atomic mass. The contribution of each broadening mechanism is determined by the relative widths of transitions of different atomic masses.

The interesting outcome was that a mixed-$b$ model generally requires fewer components and also avoids spurious double-components in line centers \citep[see][]{Lee2020}. Further, once mixed-$b$ models have been derived, it becomes apparent that temperature parameters are genuinely required by the data. 
The weighted average temperature is $T=9.12\pm1.08\times\SI{e3}{\kelvin}$. The temperature is poorly estimated for some components (due to line blending and/or weak lines). Similar results are obtained for the ThAr-calibrated spectrum (not reported).
The normalised $\chi^2$ of temperature measurements from 62 velocity components is 1.042, so the data appear to be consistent with a single temperature applying to all components. 

\subsection{Subdividing the absorption complex -- 5 regions}
\label{sec:5alpha}
We initially divide the system into five regions, I to V (coloured regions in Figure~\ref{fig:spec}). Partitioning occurs where the normalised continuum recovers to unity.  There is slight overlap between continuum regions in order to optimise continuum estimates. This partitioning has the benefit of simplifying computations and providing independent $\alpha$ measurements, whilst avoiding any potential bias that could occur if unidentified line blending corrupts part of the data.

The five $\daa$ measurements derived from splitting the absorption system into regions are tabulated in \tabref{tab:5alpha}. The quoted uncertainties are derived from the covariance matrix at the best solution. Other relevant statistical information, i.e.\ the number of free parameters for each region, the number of metal components and their average redshift, and the reduced $\chi^2$ of the model ($\chi^2_\nu=\chi^2/\nu$, where $\nu$ is the number of degree of freedom in the model), are also given. The weighted average of $\daa$ measurements over all five regions for the LFC-calibrated spectrum is $0.94\pm1.97\times10^{-6}$. The same quantity for the ThAr-calibrated spectrum is $4.82\pm1.92\times10^{-6}$.  

For the LFC-calibrated spectrum, the $\alpha$ measurements from regions I, III, IV, and V are consistent with no variation in $\alpha$. However, region II produces the seemingly anomalous result of $\daa=17.74\pm4.30\times10^{-6}$ (i.e. a \ $4.1\sigma$ deviation from zero). For the ThAr-calibrated spectrum, regions III, IV, and V are all consistent with $\daa=0$. However, regions I and II are not. Region II produces a non-zero result that is similar to the LFC spectrum. Region I also gives a strongly positive result. We discuss ways in which such anomalies can arise in \secref{sec:manyalpha}.

\begin{table}
    \centering
    \caption{Measurements when a single $\daa$ parameter is used per region (\secref{sec:5alpha}), tabulated separately for the LFC and the ThAr models. Column 1 indicates the spectral region (see Figure~\ref{fig:spec}). Columns 2 and 3 give the number of metal components ($N_c$) and the number of free parameters ($N_p$) in each model. The average redshift of the metal components is in column 4. Columns 5 and 6 give the values of $\daa$ and their $1\sigma$ uncertainties from the best-fit covariance matrix, respectively. Both are in units $10^{-6}$. The normalised $\chi_\nu^2$ for the fit is in column 7. The lower row gives the average over all five regions for the relevant quantities.}
    \label{tab:5alpha}

    \begin{tabular}{ccccS[table-format=3.3]S[table-format=3.3]c}
        \multicolumn{7}{c}{LFC}\\
        \vspace{0.1cm}
         ID & $N_c$ & $N_p$ & $\langle z \rangle$ &  \multicolumn{1}{c}{$\frac{\Delta\alpha }{\alpha}$}& \multicolumn{1}{c}{$\sigma_{stat}$ }&
         \multicolumn{1}{c}{$\chi_{\nu}^2$} \\
               
        \hline\hline
         I   & 13 & 125 & 1.14708 &  -3.90 & 4.42  & 0.9892 \\
         II  & 6  & 68  & 1.14788 & 17.74 & 4.30  & 0.9859 \\
         III & 10 & 125 & 1.14870 & 18.45 & 15.07 & 0.9836 \\
         IV  & 7  & 89  & 1.14983 & -6.39 & 4.12  & 0.8595 \\
         V   & 26 & 267 & 1.15080 & -2.59 & 3.37  & 0.9860 \\
         \hline
         All & 62   &     & 1.14936 &  0.94 & 1.97  & \\
         \hline
    \end{tabular}

\vspace{0.5cm}

    \begin{tabular}{ccccS[table-format=3.3]S[table-format=3.3]c}
        \multicolumn{7}{c}{ThAr}\\
        \vspace{0.1cm}
         ID & $N_c$ & $N_p$ & $\langle z \rangle$ &  \multicolumn{1}{c}{{$\frac{\Delta\alpha }{\alpha}$}}& \multicolumn{1}{c}{$\sigma_{stat}$ }&
         \multicolumn{1}{c}{$\chi_{\nu}^2$} \\
               
        \hline\hline
         I   & 14 & 134 & 1.14735 & 14.68 & 4.13 & 0.9652 \\
         II  & 6  & 68  & 1.14788 & 18.03 & 4.27 & 0.9730 \\
         III & 10 & 116 & 1.14872 & 4.71 & 15.67 & 0.9662 \\
         IV  & 7  & 95  & 1.14984 & -3.04 & 4.15  & 0.8343 \\
         V   & 26 & 245 & 1.15078 & -2.89& 3.12  & 0.9868 \\
          \hline
         All &  63  &     & 1.14949 &  4.82 & 1.92  & \\
         \hline
    \end{tabular}

\end{table}
\begin{table}
    \caption{The weighted average of the 47 individual $\daa$ measurements, grouped by region. Column 1 identifies the spectral region. Column 2 gives the total number of $\daa$ measurements in the region. Column 3 gives the weighted average redshift. Column 4 and 5 give the weighted average $\daa$ and associated error on the mean in units $1\times10^{-6}$. Column 6 indicates which absorption components were LTS trimmed (see Figures in Appendices \ref{sec:mvpfit_lfc_models} and \ref{sec:mvpfit_thar_models}). Columns 7 and 8 give the weighted average $\daa$ and uncertainty after applying LTS. The lower row provides weighted averages over all 47 measurements (40 after LTS). Superscripts in Column 2 and below the table identify which absorption components were removed by LTS. Where a component lies in a group, the entire group was discarded.}
\label{tab:manyalpha}
    \begin{threeparttable}
    
    \begin{tabular}{cccS[table-format=2.2]S[table-format=2.2]cS[table-format=3.2]S[table-format=3.2]}
         \multicolumn{8}{c}{LFC}\\
         \vspace{-0.2cm}\\
         ID & $N$  & $\langle z \rangle$ &
         \multicolumn{1}{c}{$\langle\frac{\Delta\alpha}{\alpha}\rangle$ }& \multicolumn{1}{c}{$\sigma_{stat}$ }&
         LTS &
         \multicolumn{1}{c}{$\langle\frac{\Delta\alpha}{\alpha}\rangle_{\rm LTS}$}&
         \multicolumn{1}{c}{$\sigma_{stat}^{\rm LTS}$}\\
         \hline\hline
         I   & 13 & 1.14707 & -5.40 &  5.47 & ap,au & -5.32 & 5.53 \\
         II  & 6  & 1.14784 & 14.17 &  4.71 & ac & 24.45 & 9.51 \\
         III & 10 & 1.14877 & 19.51 & 12.48 & au & 11.53 & 12.58 \\
         IV  & 4\tnote{a}& 1.14984 & -6.77 &  4.25 & aj & -0.50 & 5.00 \\
         V   & 14\tnote{b} & 1.15077 & -2.98 & 3.45 & ab,ar & -2.38 & 3.48  \\
         \hline
         All & 47  & 1.14943 & -0.18 & 2.11 &   & -0.27 & 2.41 \\
         \hline
         
    \end{tabular}

\vspace{0.3cm}
    \begin{tabular}{ccS[table-format=1.5]S[table-format=2.2]S[table-format=2.2]cS[table-format=3.2]S}
        
         \multicolumn{8}{c}{ThAr}\\
         \vspace{-0.2cm}\\
         ID & $N$  & $ \langle z \rangle$ &
         \multicolumn{1}{c}{$\langle\frac{\Delta\alpha}{\alpha}\rangle$ }& \multicolumn{1}{c}{$\sigma_{stat}$ }&
         LTS &
         \multicolumn{1}{c}{$\langle\frac{\Delta\alpha}{\alpha}\rangle_{\rm LTS}$}&
         \multicolumn{1}{c}{$\sigma_{stat}^{\rm LTS}$}\\
               
        \hline\hline
         I   & 13\tnote{c} & 1.14732 & 21.22 &  8.51 & aw,ar & 16.47 & 8.55 \\
         II  & 6  & 1.14784 & 13.72 &  4.69  & ac & 24.67 & 10.23  \\
         III & 10 & 1.14871 &  0.72 & 12.60 & al & 11.92 & 13.51  \\
         IV  & 4\tnote{d}  & 1.14984 & -3.08 &  3.89 & ae & -5.52 & 7.02 \\
         V   & 14\tnote{e} & 1.15078 & -4.10 &  2.87 & as,ax & -3.70 & 2.90 \\
         \hline
         All & 47 & 1.14957 & 0.88 & 1.99 &  & -0.15 & 2.44  \\
         \hline
    \end{tabular}
    \begin{tablenotes}
        \item [a] Grouped: (ag,am,aa,al)
        \item [b] Grouped: (ab,aj,al,bm), (bs,bi,ah,ac), (ak,ao), (bq,bc), (ag,an), (ai,as,am), (au,bo)
        \item [c] Grouped: (at,aa)
        \item [d] Grouped: (ag,ak,aa,am)
        \item [e] Grouped: (as,bd,ah,bo), (at,aq), (ao,ay,ab), (ak,bg), (an,bq), (ag,ad), (ai,bc,bi), (ap,bm)
    \end{tablenotes}
    \end{threeparttable}
\end{table}
\subsection{Further subdivision -- 47 measurements of $\alpha$}
\label{sec:manyalpha}

Instead of dividing the complete absorption complex into five segments and obtaining five measurements of $\daa$, we can instead solve for the best fit model using one free $\daa$ parameter for each individual absorption component in the complex. Doing so provides considerably more detail and can identify any $\daa$ outliers that might ``corrupt'' an $\alpha$ measurement derived over a whole region or complex. The cost is obviously that the number of free parameters is increased.

To do this we accept the best-fit models for each of the 5 regions and use these parameters as a starting point. However, additional parameters are included to allow $\alpha$ to vary independently for each velocity component. Optimisation is done using \vpfit. In other words, we do not recommence the entire \mvpfit~fitting process from scratch. The whole absorption complex (i.e. all five regions illustrated in \figref{fig:spec}) comprises a total of 62 velocity components for the LFC-calibrated spectrum (63 for the ThAr-calibrated). An initial trial fit showed that some badly-blended (and/or weak) velocity components provided only very poor constraints. In those cases we grouped components on small scales, resulting in a total of 47 individual measurements of $\daa$.

\begin{figure*}
    \centering
    \includegraphics[width=\textwidth]{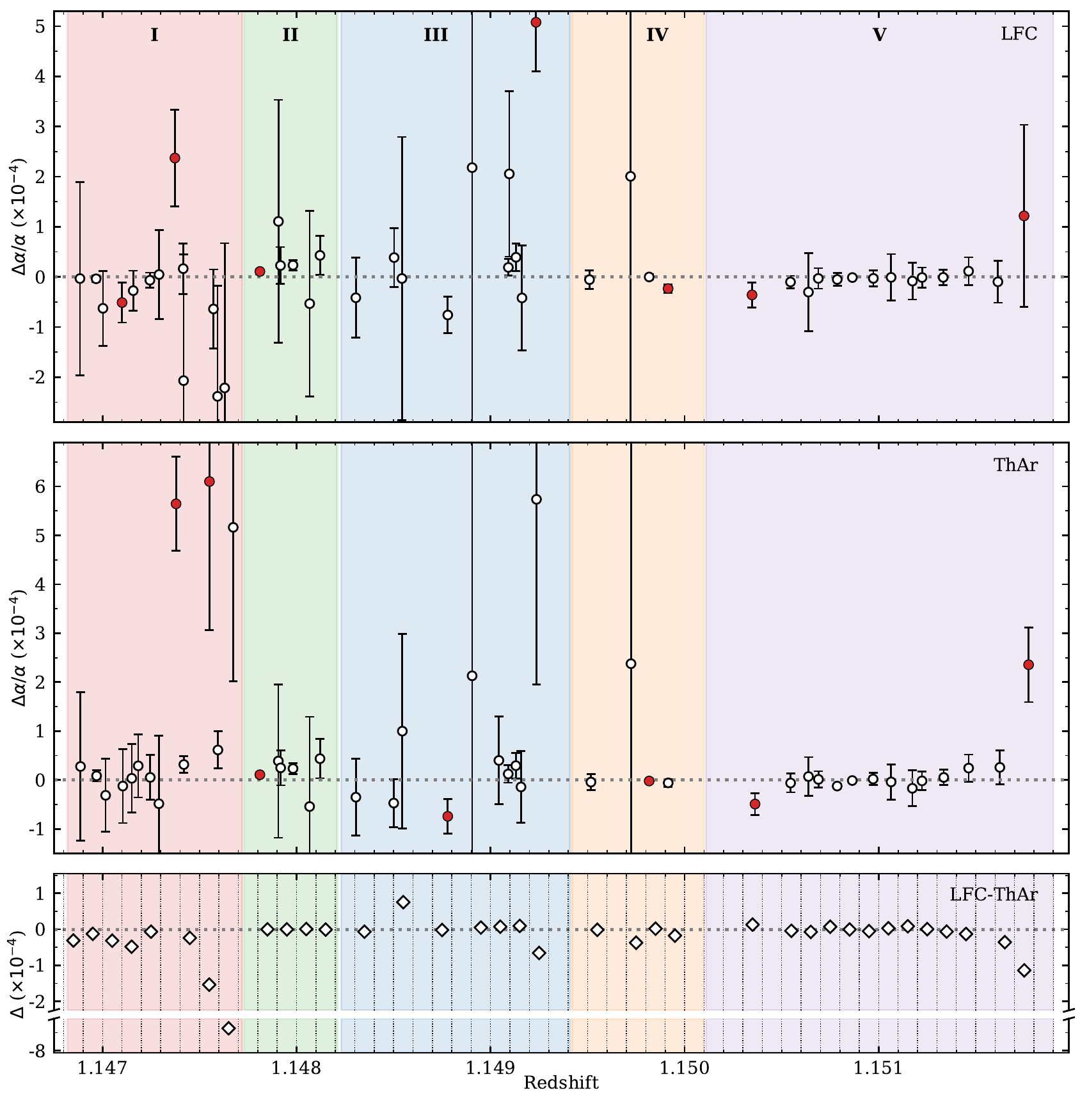}
    \caption{The 47 $\daa$ measurements. 
    Some components were grouped together (\secref{sec:manyalpha}). Measurements from the LFC-calibrated spectrum are shown in the top panel. ThAr results are shown in the middle panel. The lower panel shows the difference between the LFC and the ThAr measurements after averaging in bins of $\Delta z = 1\times 10^{-4}$ (vertical dotted lines). Weighted averages of points in each region are tabulated in \tabref{tab:manyalpha}. Filled red circles indicate measurements removed by least trimmed squares. The weighted averages for each region, after discarding those points, are also given in \tabref{tab:manyalpha}. The data illustrated in the top two panels are available as online supplementary material.}
    \label{fig:alpha_main}
\end{figure*}

The 47 $\daa$ measurements obtained this way are shown in \figref{fig:alpha_main} for both the LFC (top panel) and the ThAr (middle panel) models. The weighted average $\daa$ across each of the five regions is tabulated in Table \ref{tab:manyalpha}, together with their statistical uncertainties. The results are in good agreement with the results obtained previously, i.e.\ the weighted average of $\daa$ measurements within each region falls within $1\sigma$ of the results in \secref{sec:5alpha}. 

Unlike the results obtained in \secref{sec:5alpha}, the weighted average over the 47 measurements for the LFC-calibrated spectrum ($\daa=-0.18\pm2.11\times 10^{-6}$) and for the ThAr-calibrated spectrum ($\daa=0.88\pm1.99\times 10^{-6}$) are consistent with each other. As expected, given the large number of free model parameters, there is generally a slight increase in the $\daa$ error estimates (compare Tables \ref{tab:5alpha} and \ref{tab:manyalpha}).

\subsection{Consistency between 47 $\alpha$ measurements}
\label{sec:consistency}

We now explore the differences between the LFC and ThAr models in more detail. We bin the individual $\daa$ measurements in redshift bins $\Delta z = 1\times10^{-4}$ (the 47 measurements fall into 37 bins) and calculate the weighted average in each. Their differences (LFC minus ThAr) are illustrated in the lower panel of \figref{fig:alpha_main}. The LFC and ThAr measurements agree well everywhere except in region I. Most of the LFC-ThAr differences in region I are located around $-2\times10^{-5}$, with two bins (at the high-redshift end) at more negative values ($-1.8$ and $-7.6\times10^{-4}$). The top two panels of \figref{fig:alpha_main} suggest this is caused by velocity structure differences between the LFC and ThAr models. Discarding the points in these two bins and taking the weighted average of the remaining points in region I, we get $-4.90\pm5.49\times 10^{-6}$ for LFC and $15.68\pm8.54\times10^{-6}$ for ThAr, i.e\ the two remain inconsistent. 

The most significant deviation from zero occurs in region II ($\approx 3\sigma$). Measurements from the LFC-calibrated and the ThAr-calibrated spectrum are in excellent agreement in this region.

\begin{figure}
    \centering
    \includegraphics[width=\columnwidth]{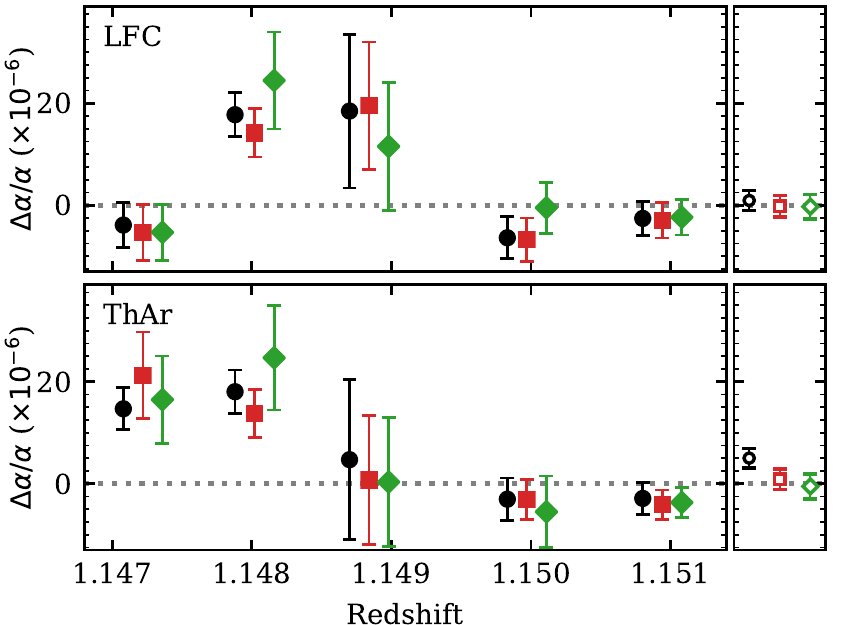}
    \caption{Comparing the different approaches to measuring $\alpha$ from the LFC-calibrated spectrum (top) and the ThAr-calibrated spectrum (bottom). Measurements from \secref{sec:5alpha} are shown as black points. The weighted average of the 47 measurements, derived in \secref{sec:manyalpha}, are shown as red squares. The green diamonds are after outlier removal. Points were offset along the $x$-axis for clarity. Panels on the right show the weighted average over the entire absorption complex.}
    \label{fig:alpha_comparison}
\end{figure} 

A substantial advantage of deriving $\daa$ measurements for individual absorbing components (or small groups) is that it may help to identify and filter out any possible rogue measurements. The least trimmed squares (LTS) method is frequently used for this and provides a more robust estimate of the mean. We thus apply LTS here, discarding 15\% of the data in each region. The discarded components are listed in column 6 of \tabref{tab:manyalpha}. The weighted averages for each region and the entire sample after LTS trimming are tabulated in columns 7 and 8 of the same table. Interestingly, the consequence of removing the most outlying measurement in region II was to move the region's average towards more positive values (but doubling the error and therefore making the result less significant).

The measured scatter (i.e.\ the empirical standard deviation) of the remaining 40 LFC-calibrated measurements is $\sigma=9\times10^{-5}$. This is slightly smaller than the average error on individual measurements, $\langle\sigma^{stat}\rangle=14\times10^{-5}$. The scatter of the ThAr-calibrated measurements is $\sigma=13\times10^{-5}$ (the same as the average error).

\subsection{Consistency with previous studies}
Recently, a detailed study of this same absorption complex was carried out \citep{Kotus2017} using spectra from the UVES spectrograph \citep{Dekker2000} on the Very Large Telescope. Those spectra are higher S/N although the spectral resolution is lower. Comparing with those results reveals good consistency. In this paper we split the data into five distinct regions whereas \citet{Kotus2017} use three. However, combining our regions \{I+II\} (``left'') and \{III+IV\} (``centre'') enables the comparison (our region V corresponds to Kotu{\v s}' ``right'' region). Using the $\daa$ results from the LTS trimmed sample in our Table~\ref{tab:manyalpha} (LFC-calibrated) and Kotu{\v s}' table 4, and combining all random and systematic errors appropriately, the $\daa$ solution differ by 1.15, 1.02, and 0.92$\sigma$ respectively, left to right. 

Prior to our study, \citet{Kotus2017} was the most detailed study.  However, several prior analyses also exist \citep{Quast2004,Levshakov2005,Levshakov2006,Chand2006,Molaro2008}. All produced results consistent with no change in $\alpha$, with somewhat larger uncertainties than derived from our analysis or that of \citet{Kotus2017}.

\section{Results}
\label{sec:results}
Figure \ref{fig:alpha_comparison} shows the $\daa$ measurements tabulated in Tables \ref{tab:5alpha} and \ref{tab:manyalpha}. The two large panels show the results from the LFC-calibrated (top panel) and the ThAr-calibrated spectrum (bottom panel). The weighted average for the entire $z\approx1.15$ absorption complex are plotted in the small panels to the right of the main panels. Our main results are summarised as follows:
\begin{enumerate}
    \item In the analysis presented in \secref{sec:5alpha}, we obtain five $\daa$ measurement (one per region) from the LFC-calibrated and from the ThAr-calibrated spectrum independently. For the LFC-calibrated spectrum, the average over the five regions is consistent with no variation in $\alpha$. Applying the same methods to the ThAr calibrated spectrum, we obtain a $2.5\sigma$ deviation from zero. These results are tabulated in \tabref{tab:5alpha} and plotted as black points in \figref{fig:alpha_comparison}.
    
    \item Including $\daa$ as a free parameter for 47 individual absorption components (or appropriately grouped components) allows us to identify regions of data significantly affecting the overall measurement (\secref{sec:manyalpha}). These results, tabulated in \tabref{tab:manyalpha} and plotted as red squares on \figref{fig:alpha_comparison}, are in excellent agreement with the results from \secref{sec:5alpha}. 
    
    \item To explore robustness, we apply LTS, removing 15\% of the sample in each region, obtaining the results plotted as green diamonds in \figref{fig:alpha_comparison}. This reduces the total number of measurements to 40. The weighted average over the 40 measurements is consistent with zero for both calibrations: $\daa=-0.27\pm2.41\times 10^{-6}$ (LFC) and $\daa=-0.15\pm2.44\times 10^{-6}$ (ThAr). 
    
    \item When using the approach in \secref{sec:5alpha}, the wavelength scale distortions imparted by the ThAr calibration methods (see \figref{fig:calibration_comparison}) have a small, but measurable, effect. The same distortions appear to have no effect on the measurements in \secref{sec:manyalpha}.
    
    \item The HE0515$-$4414 absorption complex modelled spans approximately 700 km/s. If this system represents a line of sight through a cluster of order 1 Mpc across, we can place an upper limit, for the first time, on small-scale $\alpha$ variations, using the empirical scatter in the 40 $\daa$ measurements. The upper limit on small-scale $\alpha$ variations over scale-lengths $\approx 25$ kpc, is $\approx 9 \times 10^{-5}$ (\secref{sec:consistency}).

    \item Averaged over all absorption components, we derive a gas temperature of $T=9.12\pm1.08\times\SI{e3}{\kelvin}$ (\secref{sec:temperature}). This value is in agreement with the results from \citet{Carswell2012} who found $T=12\pm3\times\SI{e3}{\kelvin}$ in a quasar absorption system at $z_{abs}=2.076$. As seen, the new data presented in this paper provide a more stringent constraint and also suggest that all individual absorption components are consistent with a single gas temperature.
    \end{enumerate}
    
\section{Discussion}
\label{sec:discussion}
In this work we have analysed the first LFC-calibrated quasar spectrum. Quasar spectra of similar quality to the one presented here will be routinely produced by the new ESPRESSO spectrograph installed on the Very Large Telescope \citep{Pepe2014} and by the future HIRES instrument on the Extremely Large Telescope \citep{Maiolino2013}. Both of these instruments have LFCs for wavelength calibration. We hope that the new methods introduced in this paper wil be beneficial in analysing future observations.

We have demonstrated that careful modelling procedures play a crucial part in the analysis of high resolution spectroscopic data. Tools such as \mvpfit~eliminate any potential human bias and yield optimal models of the data in a reproducible and objective manner.

Choosing the correct line broadening mechanism, i.e.\ including temperature parameters for individual components, is important. Models produced assuming an incorrect broadening mechanism tend to generate artificial close blends of absorption lines. Modelling simulated data, based on the HE0515 spectrum used in this work, shows that using an incorrect broadening mechanism biases $\alpha$ measurements \citep{Lee2020}. 



Examining the scatter of the five measurements obtained from the LFC-calibrated spectrum in \secref{sec:5alpha}, we find they have $\chi^2=21.99$ ($\nu=4$). For a $\chi^2$ distribution with four degrees of freedom, the probability of observing $\chi^2$ values at least this large  is $p=0.02\%$. The five measurements are therefore highly inconsistent with each other.
Performing the same for the 47 measurements from \secref{sec:manyalpha}, we find the LFC-calibrated measurements have a $\chi^2=72.86$ ($\nu=46$, $p=0.7\%$). 
We assume this is not caused by small-scale spatial variations in $\alpha$ across the redshift range covered by the absorption system. After LTS trimming, the scatter in the remaining 40 measurements are consistent with their individually estimated errors ($\chi^2=23.76$, $\nu=39$, $p=97\%$). Similar results are obtained for the ThAr spectrum.

The analysis presented here is based on the assumption of solar relative isotopic abundances. Significant deviations from solar values translate to large shifts in $\daa$ \citep{Webb1999, Ashenfelter2004PRL, Ashenfelter2004ApJ, Fenner2005, Berengut2012, Webb2014}. Very approximately, when simultaneously modelling \ion{Mg}{ii} and \ion{Fe}{ii}, the measured $\daa$ may shift towards negative values by as much as $\approx5 \times 10^{-6}$ for 100\% \atom{Mg}{24} and by the same amount in the positive direction for 100\% \atom{Mg}{25+26}. A discussion as to the validity of the solar isotopic assumption is deferred to a subsequent paper.

For these particular observations, LFC calibration methods have not yielded significantly different $\daa$ measurements compared to the ThAr methods. The probable reason for this lies in the fact that we have combined a large number of individual exposures to form a final co-added spectrum. Due to different barycentric velocities for each observation, the position of relevant transitions falls differently with respect to the complicated distortion pattern each time, effectively smearing it out. This is less likely to occur for more efficient spectrographs, such as ESPRESSO and HIRES.

\section*{Data availability}
We provide as online supplementary material: (i) the LFC and ThAr wavelength calibrated co-added spectra, (ii) the final models, and (iii) all input files for \vpfit~and \mvpfit, including atomic data.

\section*{Acknowledgements}
We wish to dedicate this work to our dear colleague and friend John Barrow, who has played such an important role in the development of this subject. Based on observations collected at the European Organisation for Astronomical Research in the Southern Hemisphere under ESO programme 102.A-0697(A). We are grateful for the award of computing time for this research on the gStar and OzStar supercomputing facilities at the  Centre for Astrophysics and Supercomputing, Swinburne University of Technology. DM thanks Prashin Jethwa for useful discussions during the early stages of the analysis. JKW thanks the John Templeton Foundation, the Department of Applied Mathematics and Theoretical Physics and the Institute of Astronomy at Cambridge University for hospitality and support, and Clare Hall for a Visiting Fellowship during this work. We thank the referee for their useful comments. 




\bibliographystyle{mnras}
\bibliography{biblio,atomic} 




\appendix

\section{Atomic data}
\begin{table*}
\begin{center}
\caption{
Atomic species and transitions, with isotopic structure, used in this analysis. Terrestrial isotopic relative abundances are assumed. Column 4 ($\lambda_0$) is rest-frame wavelength. Column 5 ($f$) is oscillator strength or relative abundance (\%). The latter are from \citet{Rosman1998}. Column 6 ($\Gamma$) is the sum of the spontaneous emission rates. Column 7 ($q$) gives the sensitivity coefficients to a change in the fine structure constant $\alpha$. Citations to original measurement papers are given at the foot of the table. An atomic data compilation including the data in this table is given in \citet{Murphy2014}.
\label{tab:transitions}\vspace{-0.5em}
}
\begin{tabular}{l l c l l l l}\hline
\multicolumn{1}{c}{Ion}&
\multicolumn{1}{c}{Tran.}&
\multicolumn{1}{c}{$A$}&
\multicolumn{1}{c}{$\lambda_0$\;(\SI{}{\angstrom})}&
\multicolumn{1}{c}{$f$ or {\it \%}}&
\multicolumn{1}{c}{$\Gamma\;(\SI{e8}{\per\second})$}&
\multicolumn{1}{c}{$q\;(\SI{}{\per\centi\meter})$}\\
\hline
\hline

%
Fe{\sc \,ii}   & 2344   & 55.845    & 2344.212747(76)$^{a,b}$   &  0.114    & \SI{2.680}{}$^{c,d}$ & $ 1375^{e,f}(300)$\\
               &        & 58        & 2344.2113616$^{f}$        & 0.282\%   &                & $     ^{}     $\\
               &        & 57        & 2344.2120103$^{f}$        & 2.119\%   &                &$     ^{}     $\\
               &        & 56        & 2344.2126822$^{f}$        & 91.754\%  &                & $     ^{}     $\\
               &        & 54        & 2344.2141007$^{f}$        & 5.845\%   &                & $     ^{}     $\\
               & 2374   & 55.845    & 2374.460064(78)$^{a,b}$   & 0.03130   & \SI{3.090}{}$^{c,g}$ & $ 1625^{e,f}(100)$\\
               &        & 58        & 2374.4582998$^{f}$        & 0.282\%   &                &$     ^{}     $\\
               &        & 57        & 2374.4591258$^{f}$        & 2.119\%   &                &$     ^{}     $\\
               &        & 56        & 2374.4599813$^{f}$        & 91.754\%  &                &$     ^{}     $\\
               &        & 54        & 2374.4617873$^{f}$        & 5.845\%   &                &$     ^{}     $\\
               & 2382   & 55.845    & 2382.763995(80)$^{a,b}$   & 0.320     & \SI{3.130}{}$^{c,g}$ &$ 1505^{e,f}(100)$\\
               &        & 58        & 2382.7622294$^{f}$        & 0.282\%   &                & $     ^{}     $\\
               &        & 57        & 2382.7630560$^{f}$        & 2.119\%   &                &$     ^{}     $\\
               &        & 56        & 2382.7639122$^{f}$        & 91.754\%  &                &$     ^{}     $\\
               &        & 54        & 2382.7657196$^{f}$        & 5.845\%   &                &$     ^{}     $\\
               & 2586   & 55.845    & 2586.649312(87)$^{a,b}$   & 0.0691    & \SI{2.720}{}$^{c}$ &$ 1515^{e,f}(100)$\\
               &        & 58        & 2586.6475648$^{f}$        & 0.282\%   &                &$     ^{}     $\\
               &        & 57        & 2586.6483830$^{f}$        & 2.119\%   &                &$     ^{}     $\\
               &        & 56        & 2586.6492304$^{f}$        & 91.754\%  &                &$     ^{}     $\\
               &        & 54        & 2586.6510194$^{f}$        & 5.845\%   &                &$     ^{}     $\\
               & 2600   & 55.845    & 2600.172114(88)$^{a,g}$   & 0.239     & \SI{2.700}{}$^{c}$ &$ 1370^{e,f}(100)$\\
               &        & 58        & 2600.1703603$^{f}$        & 0.282\%   &                &$     ^{}     $\\
               &        & 57        & 2600.1711816$^{f}$        & 2.119\%   &                &$     ^{}     $\\
               &        & 56        & 2600.1720322$^{f}$        & 91.754\%  &                &$     ^{}     $\\
               &        & 54        & 2600.1738281$^{f}$        & 5.845\%   &                &$     ^{}     $\\
Mg{\sc \,i}    & 2852   & 24.3050   & 2852.962797(15)$^{}$      & 1.83      & \SI{5.000}{}$^{h,i,j,k,l}_{m,n,o,p}$ &$   90^{q,r}(10) $\\
               &        & 26        & 2852.959591(20)$^{s}$     & 11.01\%   &                &$     ^{}     $\\
               &        & 25        & 2852.961407(20)$^{s}$     & 10.00\%   &                &$     ^{}     $\\
               &        & 24        & 2852.963420(14)$^{s}$     & 78.99\%   &                &$     ^{}     $\\
Mg{\sc \,ii}   & 2796   & 24.3050   & 2796.353790(16)$^{}$      & 0.6155    & \SI{2.625}{}$^{t}$ &$  212^{u}(2)  $\\
               &        & 26        & 2796.34704565(42)$^{v}$   & 11.01\%   &                &$     ^{}     $\\
               &        & 25        & 2796.353449(50)$^{v,w,x}$ & 4.17\%    &                &$     ^{}     $\\
               &        & 25        & 2796.349030(50)$^{v,w,x}$ & 5.83\%    &                &$     ^{}     $\\
               &        & 24        & 2796.35509903(42)$^{v}$   & 78.99\%   &                &$     ^{}     $\\
               & 2803   & 24.3050   & 2803.530982(16)$^{}$      & 0.3058    & \SI{2.595}{}$^{t}$ &$  121^{u}(2)  $\\
               &        & 26        & 2803.52420938(42)$^{v}$   & 11.01\%   &                &$     ^{}     $\\
               &        & 25        & 2803.530941(50)$^{v,w,x}$ & 4.17\%    &                &$     ^{}     $\\
               &        & 25        & 2803.525985(50)$^{v,w,x}$ & 5.83\%    &                &$     ^{}     $\\
               &        & 24        & 2803.53229720(42)$^{v}$   & 78.99\%   &                &$     ^{}     $\\
Mn{\sc \,ii}   & 2576   & 54.9380   & 2576.87512(11)$^{a,b,y}$  & 0.361     & \SI{2.820}{}$^{z,\textsc{a},\textsc{b},\textsc{c}}$ &$ 1276^{\textsc{d}}(150)$\\
               &        & 55        & 2576.890898$^{}$          & 28.571\%  &                &$     ^{}     $\\
               &        & 55        & 2576.879368$^{}$          & 23.801\%  &                &$     ^{}     $\\
               &        & 55        & 2576.869849$^{}$          & 19.030\%  &                &$     ^{}     $\\
               &        & 55        & 2576.862494$^{}$          & 14.286\%  &                &$     ^{}     $\\
               &        & 55        & 2576.856181$^{}$          & 14.312\%  &                &$     ^{}     $\\
               & 2594   & 54.9380   & 2594.49643(11)$^{a,b,y}$  & 0.280     & \SI{2.780}{}$^{z,\textsc{a},\textsc{b},\textsc{c}}$ &$ 1030^{\textsc{d}}(150)$\\
               &        & 55        & 2594.512068$^{}$          & 28.579\%  &                &$     ^{}     $\\
               &        & 55        & 2594.500587$^{}$          & 23.841\%  &                &$     ^{}     $\\
               &        & 55        & 2594.491191$^{}$          & 19.078\%  &                &$     ^{}     $\\
               &        & 55        & 2594.483901$^{}$          & 14.289\%  &                &$     ^{}     $\\
               &        & 55        & 2594.477608$^{}$          & 14.213\%  &                &$     ^{}     $\\
               & 2606   & 54.9380   & 2606.45877(11)$^{a,b,y}$  & 0.198     & \SI{2.270}{}$^{z,\textsc{a},\textsc{b},\textsc{c}}$ &$  869^{\textsc{d}}(150)$\\
               &        & 55        & 2606.478271$^{}$          & 28.563\%  &                &$     ^{}     $\\
               &        & 55        & 2606.463977$^{}$          & 23.793\%  &                &$     ^{}     $\\
               &        & 55        & 2606.452264$^{}$          & 19.052\%  &                &$     ^{}     $\\
               &        & 55        & 2606.443176$^{}$          & 14.282\%  &                &$     ^{}     $\\
               &        & 55        & 2606.435406$^{}$          & 14.310\%  &                &$     ^{}     $\\
\hline

\end{tabular}
\end{center}

\begin{flushleft} { \footnotesize
$^{a}$\citet{Aldenius2009};
$^{b}$\citet{Nave2012};
$^{c}$\citet{BBKAP91};
$^{d}$\citet{GAPJB92};
$^{e}$\citet{Dzuba2002};
$^{f}$\citet{Porsev2009};
$^{g}$\citet{SMH98};

$^{h}$\citet{L64};
$^{i}$\citet{SG66};
$^{j}$\citet{ADJS70};
$^{k}$\citet{SL71};
$^{l}$\citet{LEHM73};
$^{m}$\citet{MR73};
$^{n}$\citet{KM78};
$^{o}$\citet{LLMVJ80};
$^{p}$\citet{LS93};
$^{q}$\citet{Berengut2005};
$^{r}$\citet{Savukov2008};
$^{s}$\citet{Salumbides2006};
$^{t}$\citet{ALP89};
$^{u}$\citet{Dzuba2007};
$^{v}$\citet{Batteiger2009};
$^{w}$\citet{Itano1981};
$^{x}$\citet{Sur2005}.
$^{y}$\citet{Blackwell-Whitehead2005};
$^{z}$\citet{KMWZ82};
$^{\textsc{a}}$\citet{PGJBAB92};
$^{\textsc{b}}$\citet{SBK95};
$^{\textsc{c}}$\citet{KG00};
$^{\textsc{d}}$\citet{Berengut2004}.
}
\end{flushleft}
\end{table*}

\section{LFC-calibrated models}
\label{sec:mvpfit_lfc_models}

\begin{figure*}
    \centering
    \includegraphics[width=\textwidth]{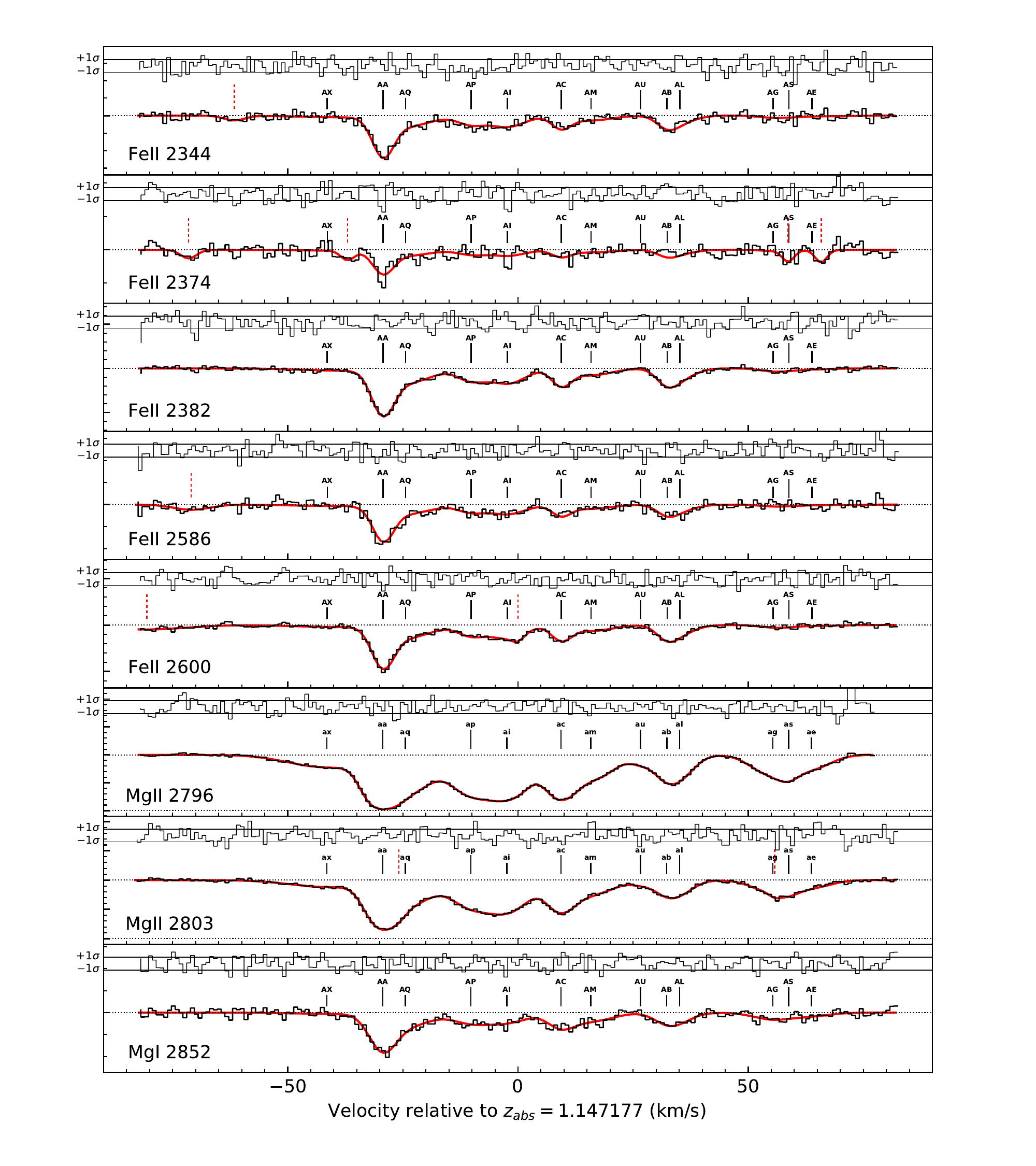}
    \caption{The black histogram shows the continuum-normalised LFC-calibrated spectrum for spectral region I. Overplotted as a continuous red line is the lowest AICc \mvpfit~model. Black labeled ticks mark the locations of absorption components in the model. The lowercase letters are associated with the transitions which provide the most information about the velocity structure \citep[i.e.\ the ``primary species'', see][]{Lee2020}. Slightly longer, dotted red ticks mark the locations of blends from unidentified species (interlopers). The black histogram above the data and the model show the normalised residuals (data-model) and the horizontal lines show the $\pm1\sigma$ levels. The dotted horizontal line corresponds to a normalised flux of unity. Major (minor) ticks on the $y$-axis label increments of 0.5 (0.1) in normalised flux.}
    \label{fig:lfc_1}
\end{figure*}
\begin{figure*}
    \centering
    \includegraphics[width=\textwidth]{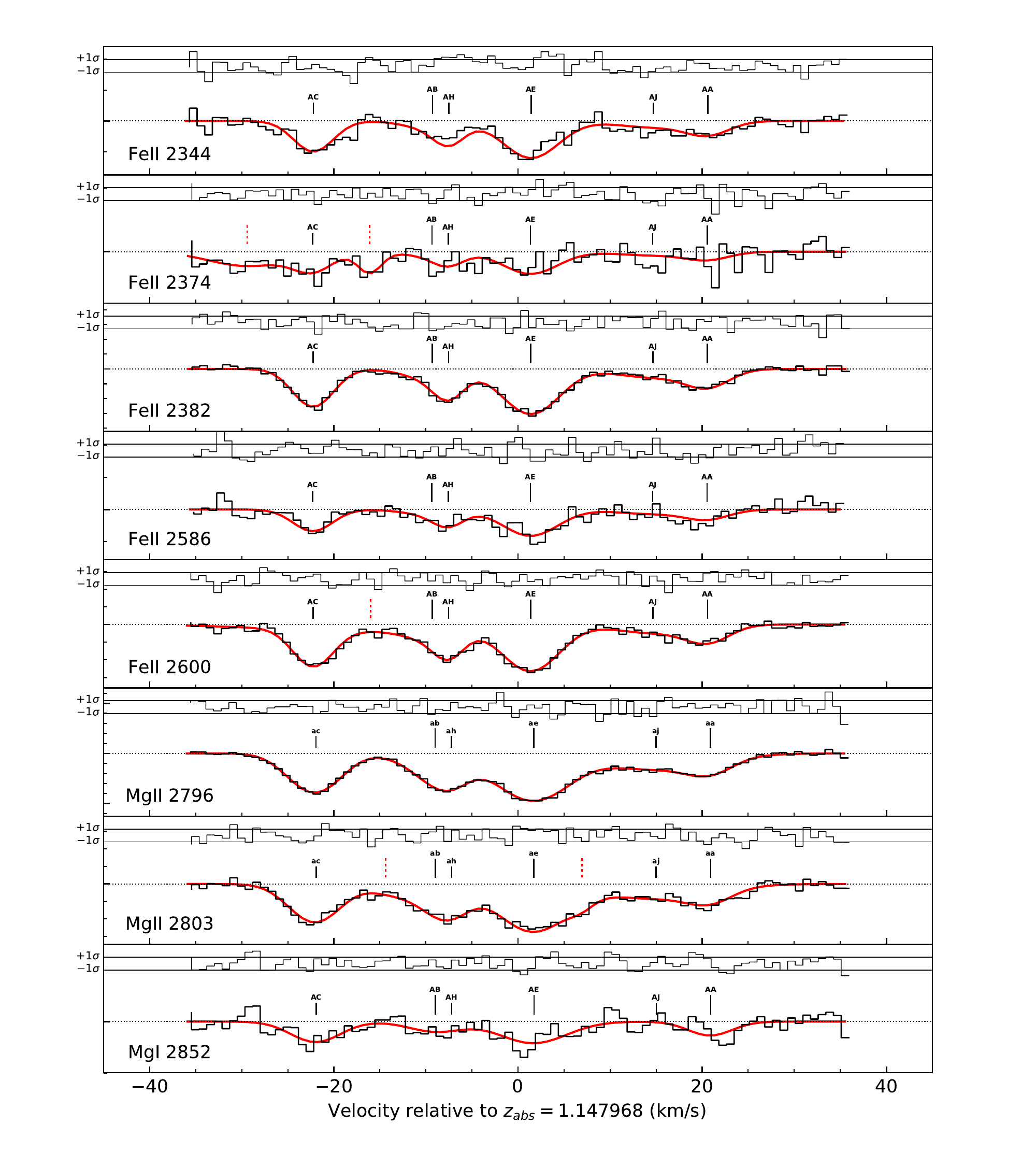}
    \caption{The same as in \figref{fig:lfc_1}, except for LFC-calibrated spectral region II.}
    \label{fig:lfc_2}
\end{figure*}
\begin{figure*}
    \centering
    \includegraphics[width=\textwidth]{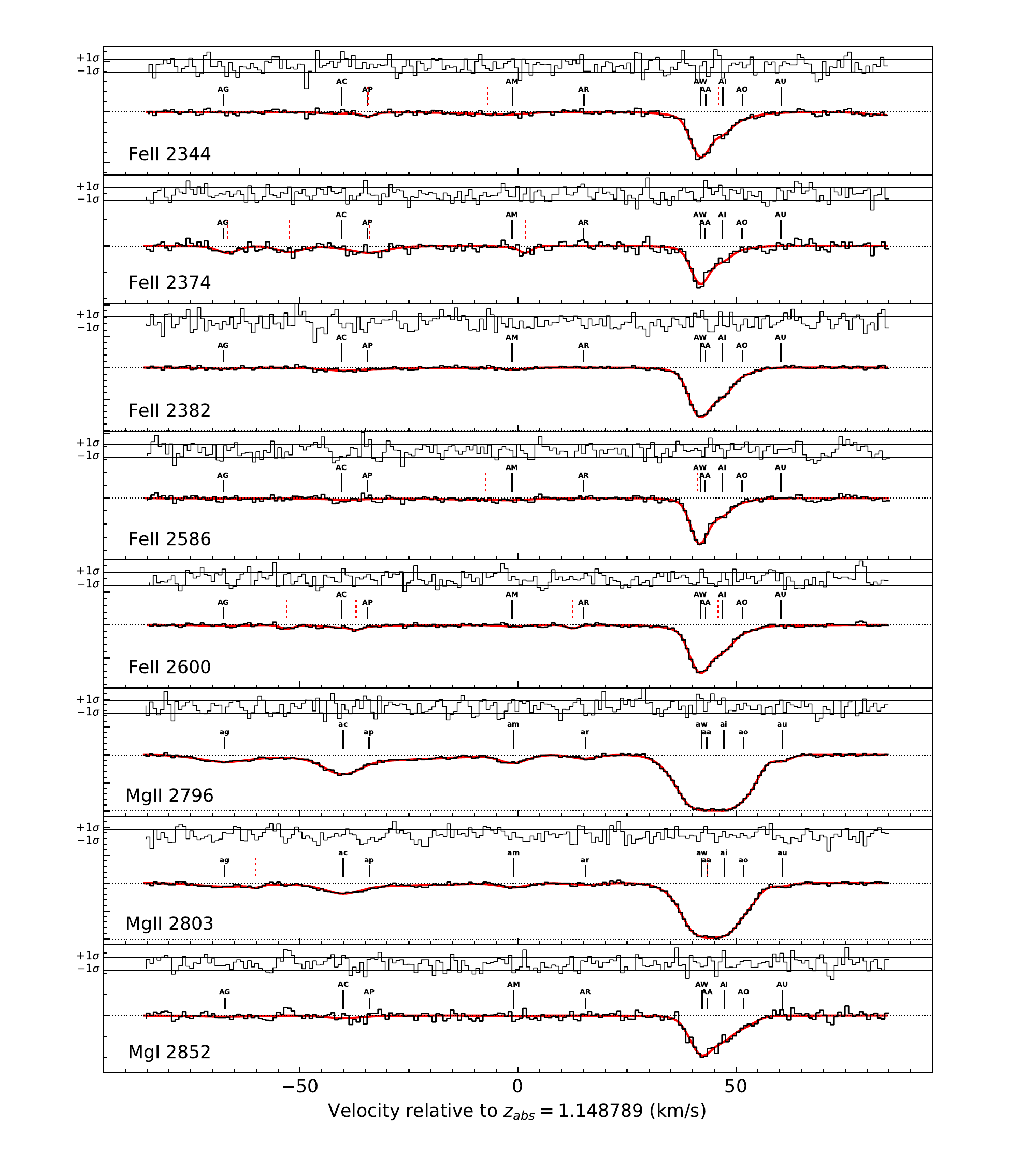}
    \caption{The same as in \figref{fig:lfc_1}, except for LFC-calibrated spectral region III.}
    \label{fig:lfc_3}
\end{figure*}
\begin{figure*}
    \centering
    \includegraphics[width=\textwidth]{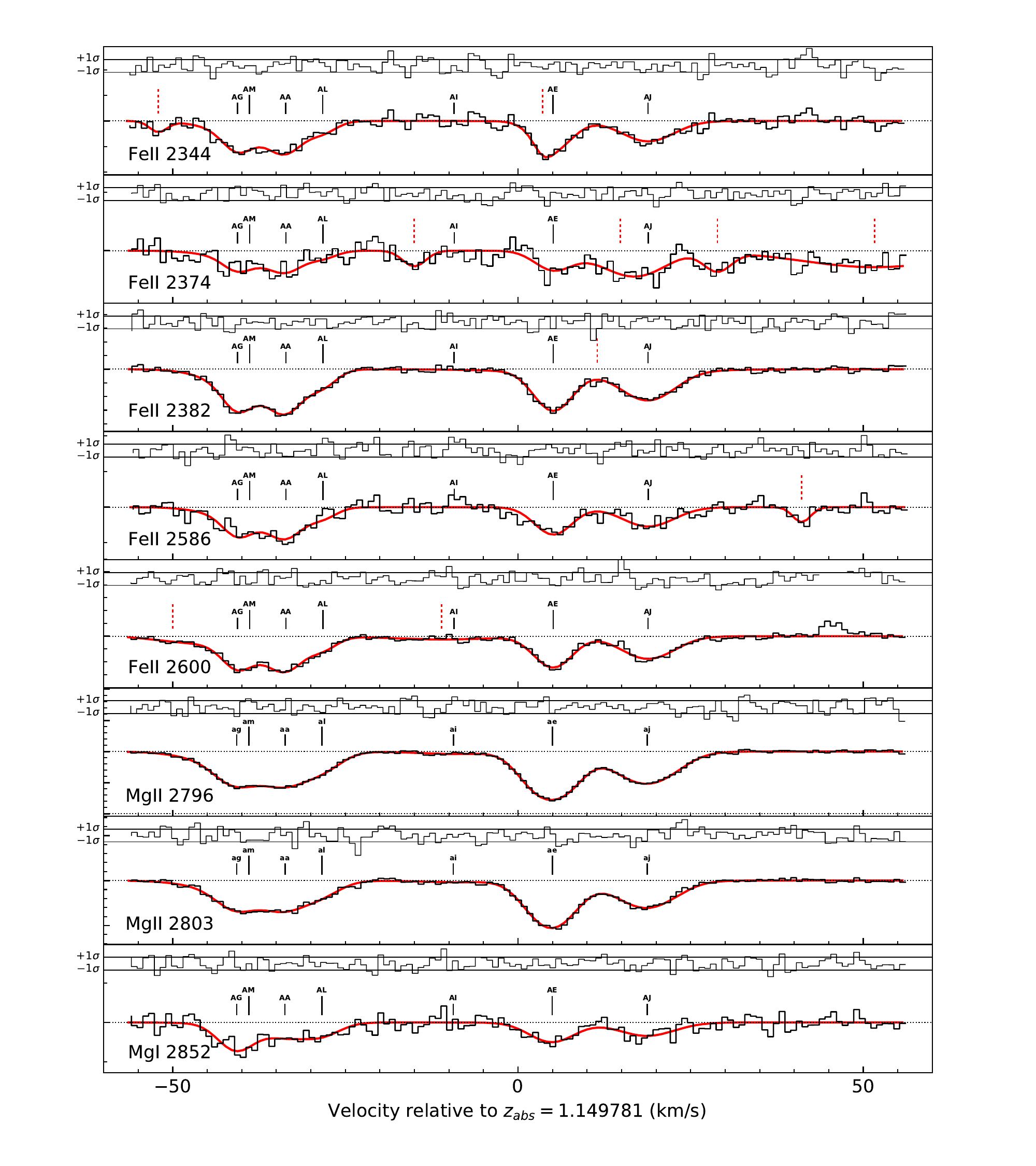}
    \caption{The same as in \figref{fig:lfc_1}, except for LFC-calibrated spectral region IV.}
    \label{fig:lfc_4}
\end{figure*}
\begin{figure*}
    \centering
    \includegraphics[width=\textwidth]{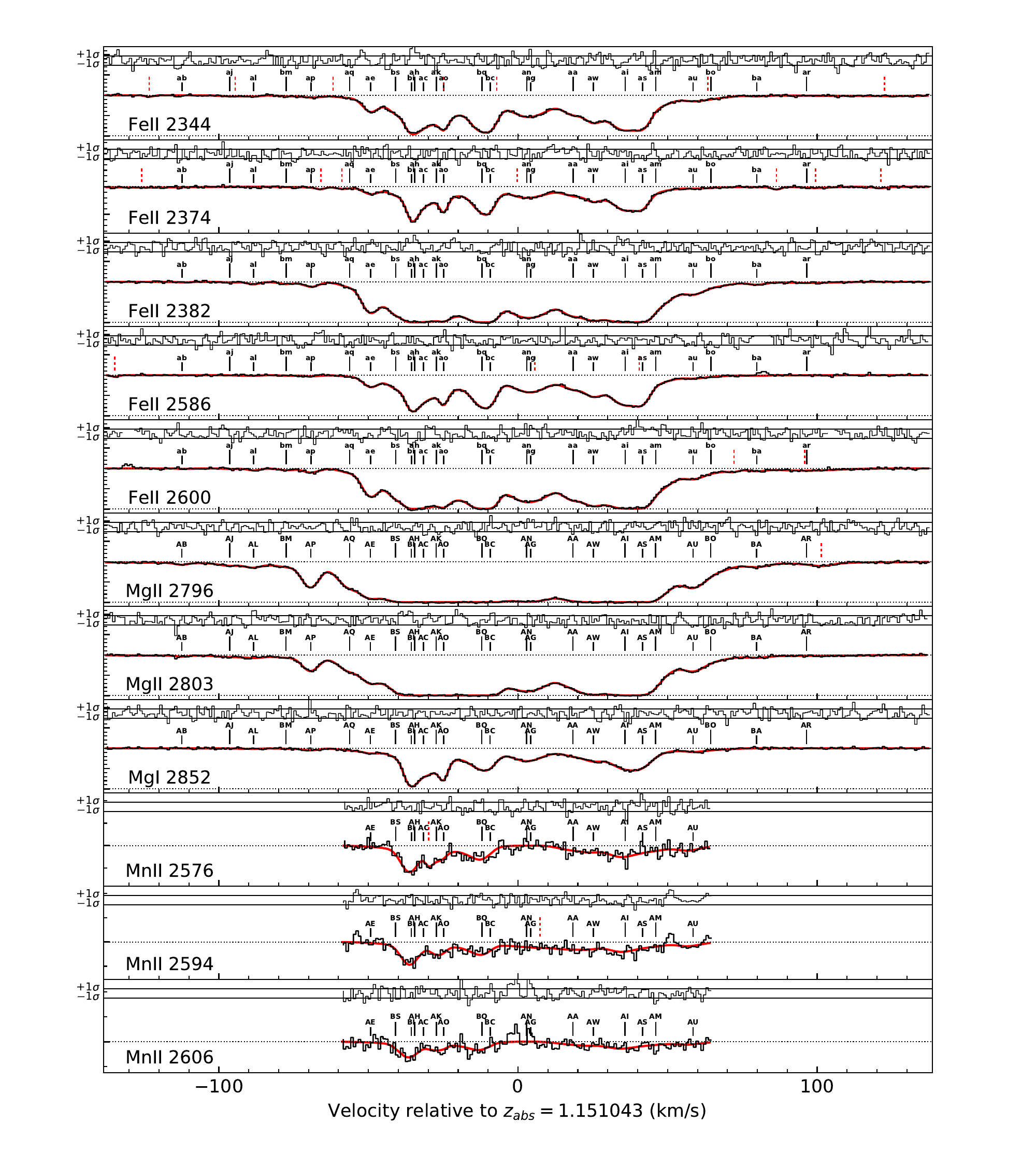}
    \caption{The same as in \figref{fig:lfc_1}, except for LFC-calibrated spectral region V.}
    \label{fig:lfc_5}
\end{figure*}

\section{ThAr-calibrated models}
\label{sec:mvpfit_thar_models}
\begin{figure*}
    \centering
    \includegraphics[width=\textwidth]{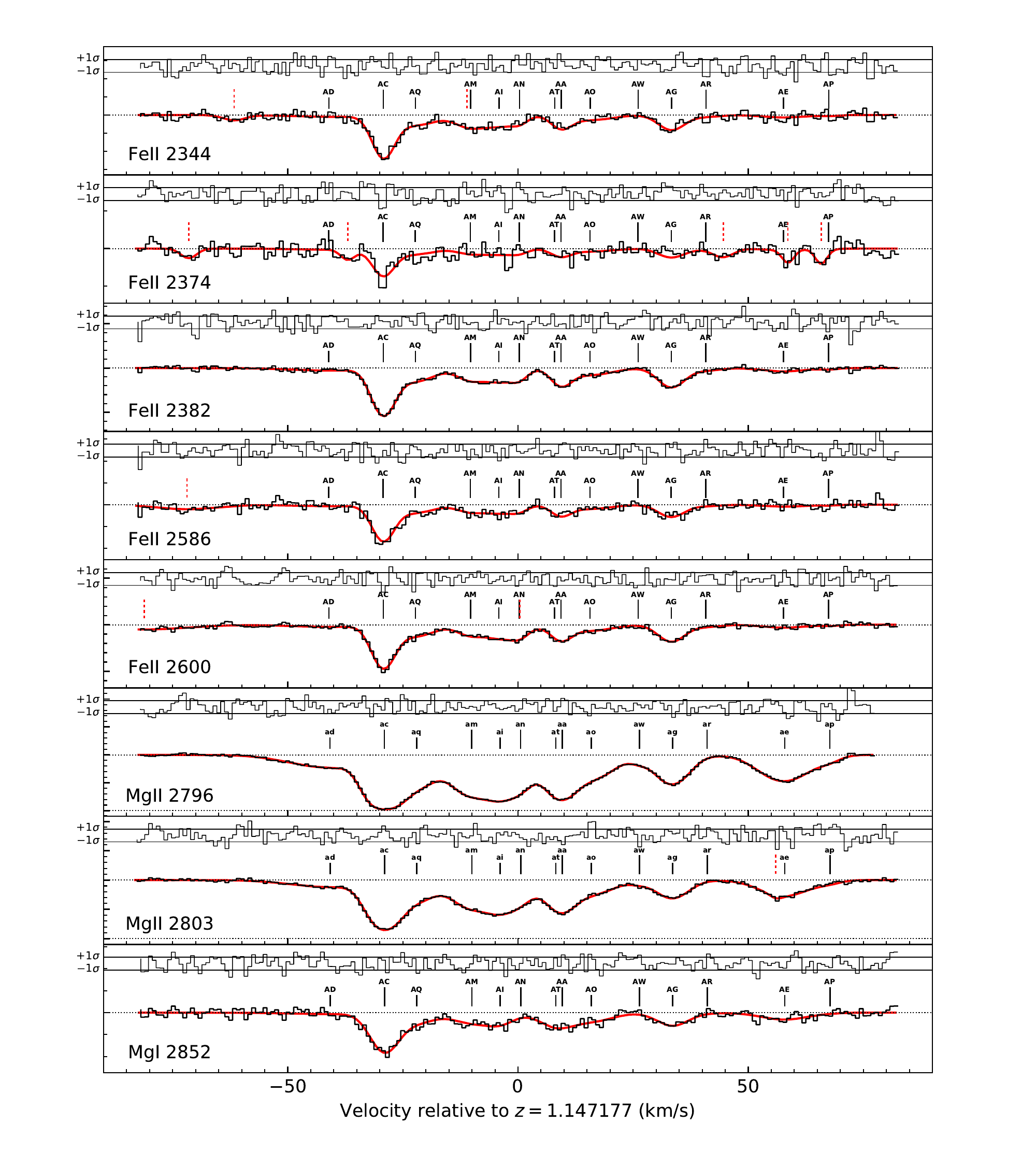}
    \caption{The same as in \figref{fig:lfc_1}, except for ThAr-calibrated spectral region I.}
    \label{fig:thar_1}
\end{figure*}
\begin{figure*}
    \centering
    \includegraphics[width=\textwidth]{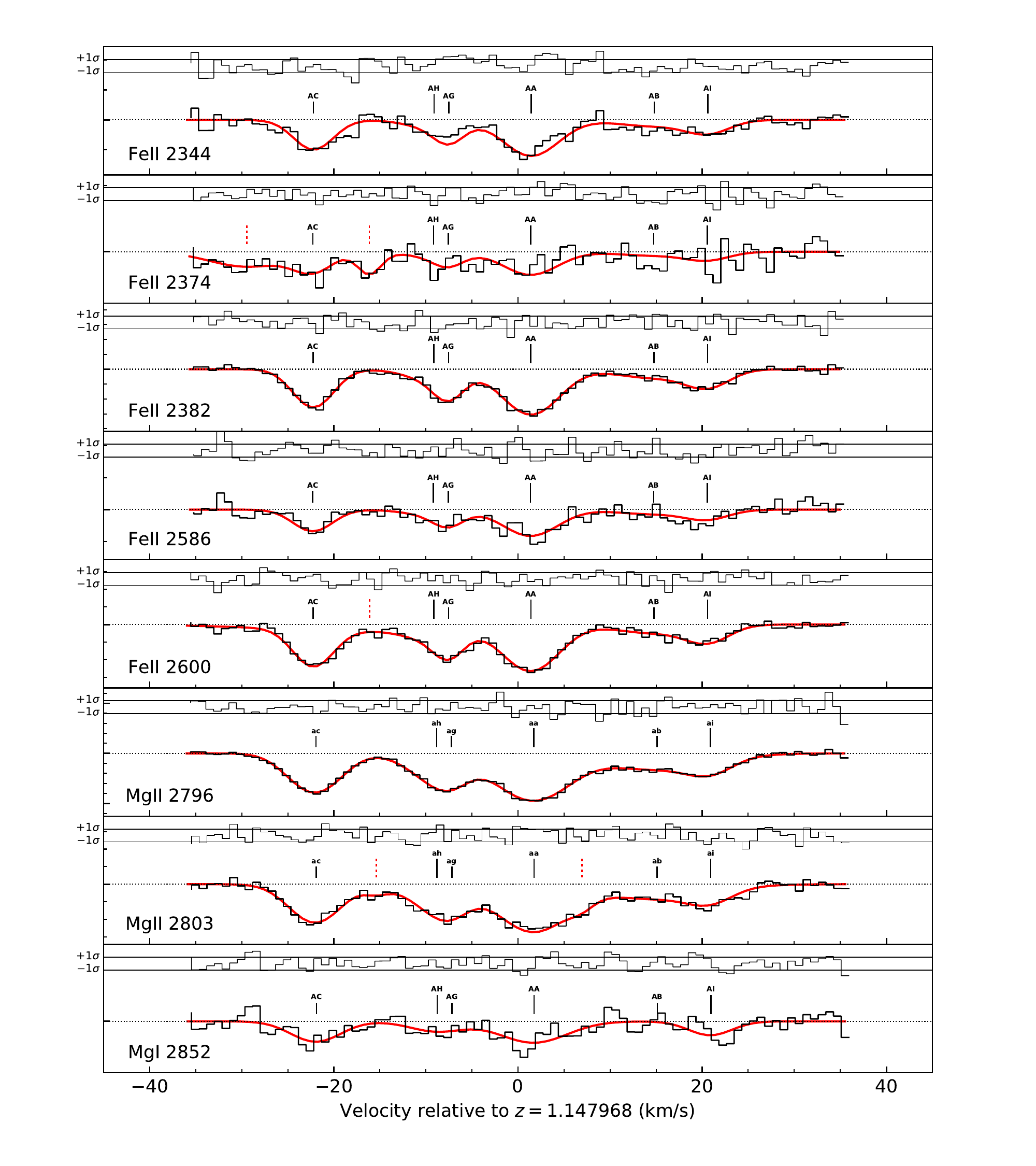}
    \caption{The same as in \figref{fig:lfc_1}, except for ThAr-calibrated spectral region II.}
    \label{fig:thar_2}
\end{figure*}
\begin{figure*}
    \centering
    \includegraphics[width=\textwidth]{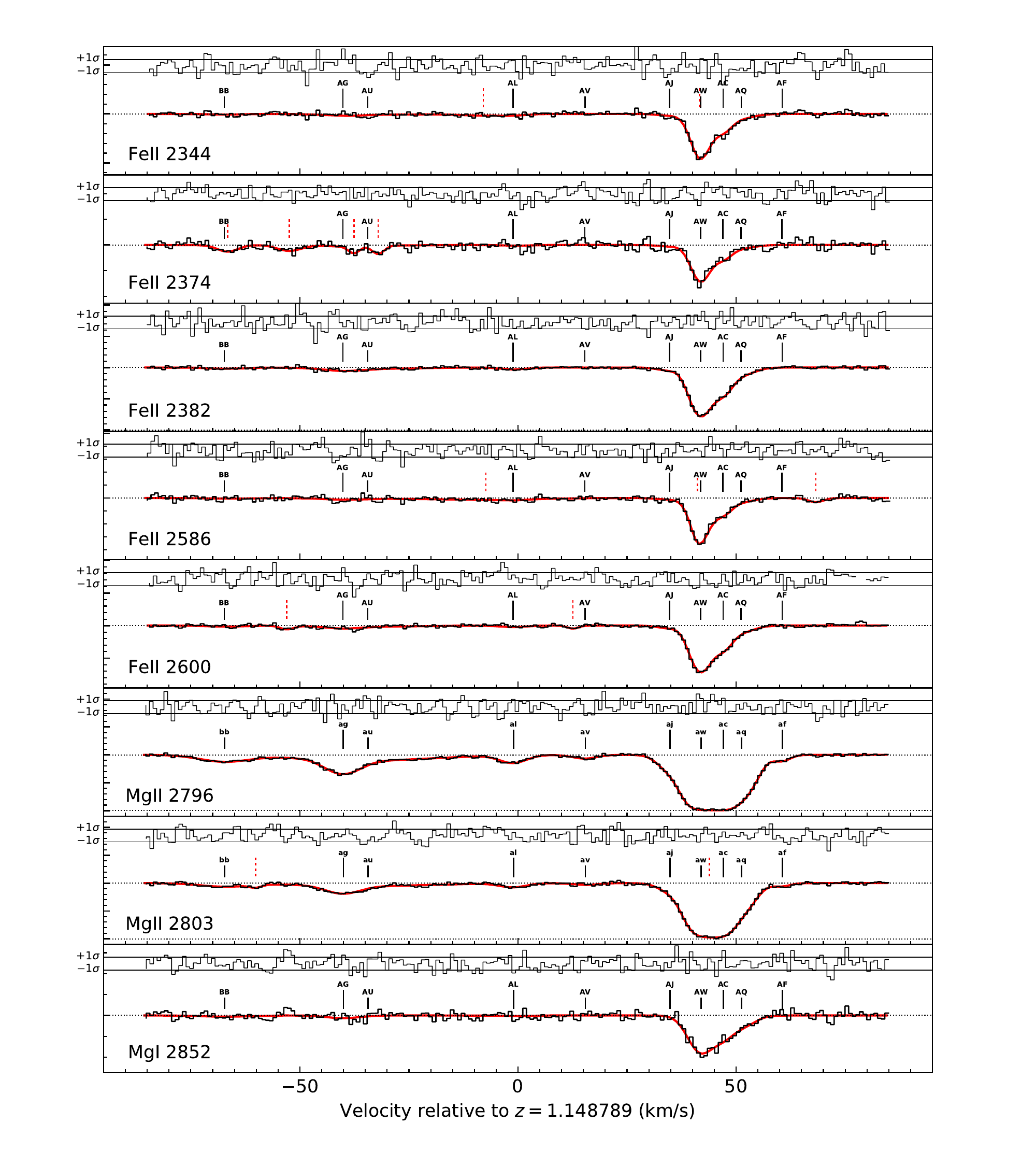}
    \caption{The same as in \figref{fig:lfc_1}, except for ThAr-calibrated spectral region III.}
    \label{fig:thar_3}
\end{figure*}
\begin{figure*}
    \centering
    \includegraphics[width=\textwidth]{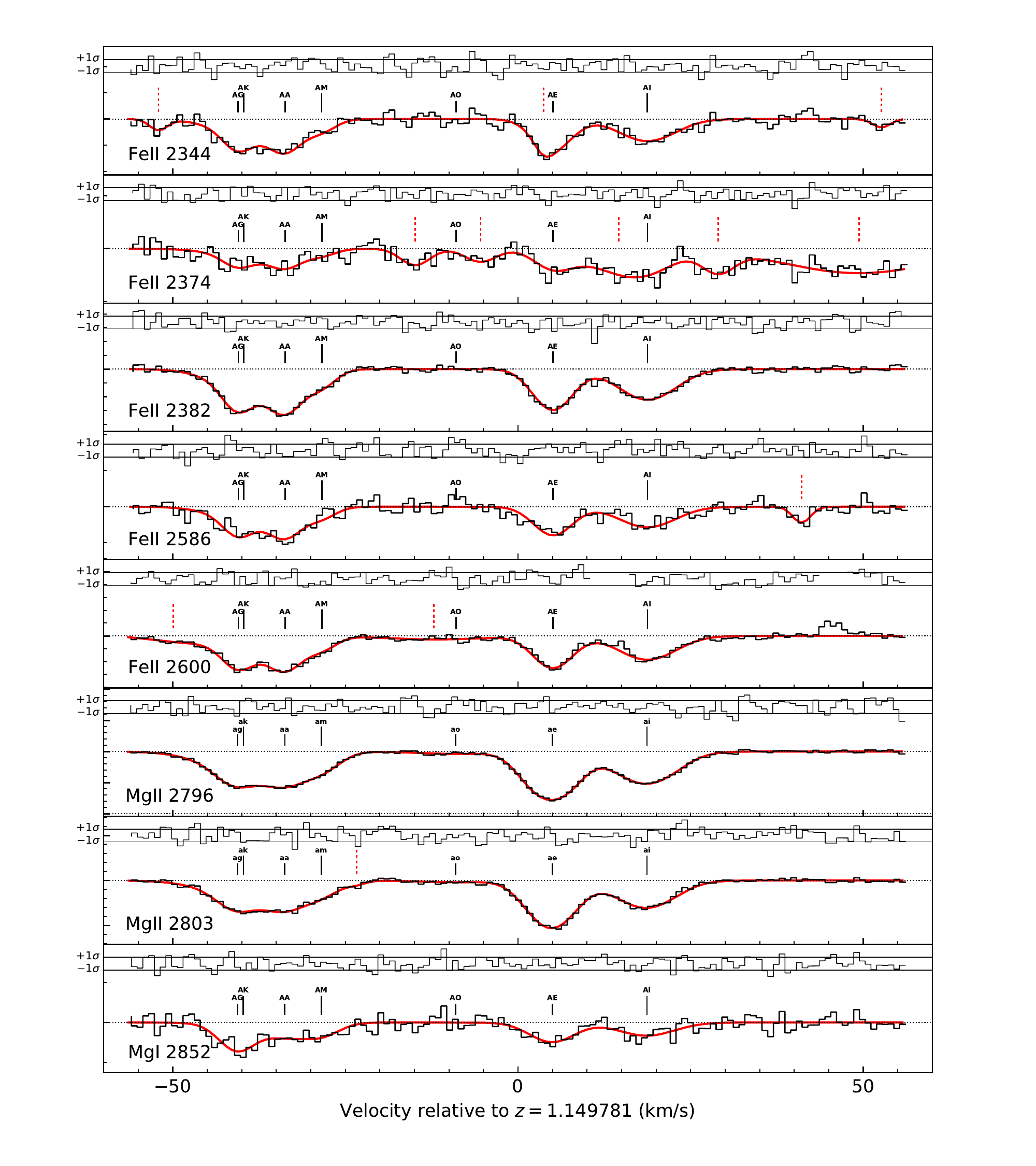}
    \caption{The same as in \figref{fig:lfc_1}, except for ThAr-calibrated spectral region IV.}
    \label{fig:thar_4}
\end{figure*}
\begin{figure*}
    \centering
    \includegraphics[width=\textwidth]{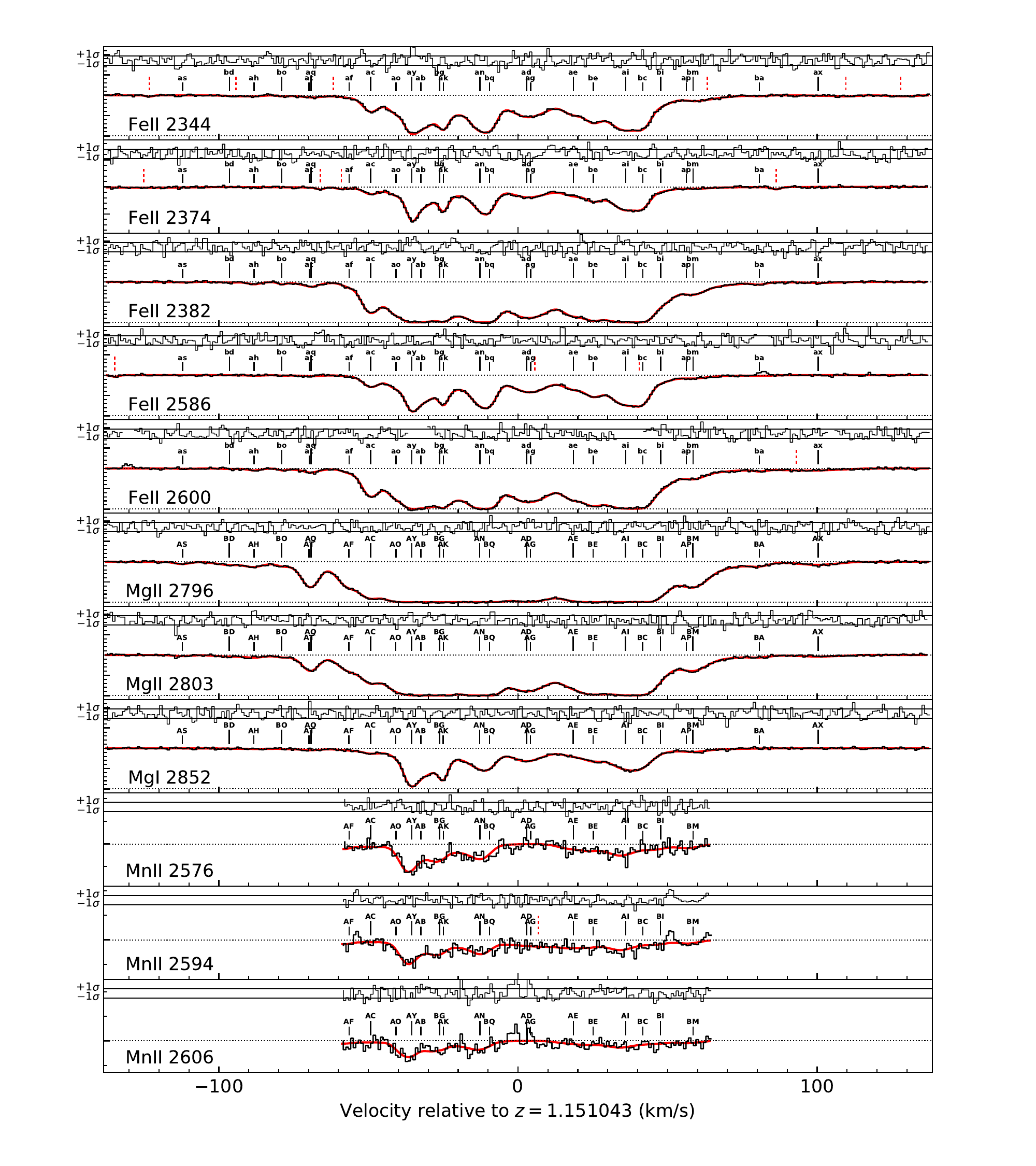}
    \caption{The same as in \figref{fig:lfc_1}, except for ThAr-calibrated spectral region V.}
    \label{fig:thar_5}
\end{figure*}


\bsp	
\label{lastpage}
\end{document}